\documentclass[11pt]{article}
\usepackage{amsmath, amssymb, graphics}

\usepackage[nosort]{cite}
\newcommand{\mathsym}[1]{{}}
\usepackage{verbatim}
\usepackage{amsfonts,euscript,amssymb,stmaryrd,braket}
\usepackage{graphics,tikz}
\usetikzlibrary{shapes.misc,arrows,decorations.markings,patterns}
\usepackage{caption}
\usepackage{subcaption}
\usepackage{slashed}
\definecolor{hyperref}{RGB}{026,028,185}
\usepackage[bookmarks=true,colorlinks=true,linkcolor=hyperref,citecolor=hyperref,urlcolor=hyperref,bookmarksnumbered]{hyperref}
 \usepackage{booktabs}
 \usepackage{multirow}
\usepackage{tabularx}
\usepackage{longtable}
\usepackage{bbold,bbding}
\usepackage{amsthm} 
\usepackage{esint}
\newcommand{\bal}{\begin{equation}\begin{aligned}}
\newcommand{\eal}{\end{aligned} \end{equation}}
\def\id{\protect{{1 \kern-.28em {\rm l}}}}

\makeatletter
\renewcommand\section{\@startsection {section}{1}{\z@}%
                                   {-3.5ex \@plus -1ex \@minus -.2ex}%
                                   {2.3ex \@plus.2ex}%
                                   {\normalfont\large\bfseries}}
\renewcommand\subsection{\@startsection{subsection}{2}{\z@}%
                                   {-3.25ex\@plus -1ex \@minus -.2ex}%
                                   {1.5ex \@plus .2ex}%
                                   {\normalfont\normalsize\bfseries}}

\makeatother

\numberwithin{equation}{section}

\usepackage[utf8]{inputenc}
\usepackage{epsfig}
\usepackage{graphicx}
\usepackage[multiple]{footmisc}
\usepackage{amssymb,amsmath}
\usepackage{fancybox,framed,tikz}
\usetikzlibrary{decorations.pathmorphing,patterns}
\usepackage{dsfont}
\usepackage{mathtools}
\usepackage{braket}
\usepackage{slashed}
\usepackage{rotating}
\usepackage{bbold,amsfonts}
\usepackage{multirow}
\usepackage{verbatim}
\usepackage{upgreek}
\newtheorem{theorem}{Theorem}

\tikzset{cross/.style={cross out, draw=black, minimum size=2*(#1-\pgflinewidth), inner sep=0pt, outer sep=0pt},
cross/.default={1pt}}

\newcommand{\be}{\begin{equation}}
\newcommand{\ee}{\end{equation}}

\renewcommand{\a}{\alpha}

\renewcommand{\d}{\delta}

\newcommand{\D}{\Delta}
\newcommand{\e}{\epsilon}

\newcommand{\m}{\mu}

\newcommand{\n}{\nu}

\usepackage{color}
\newcommand{\colb}[1]{\textcolor{blue}{#1}}

\definecolor{mypink1}{rgb}{0.958, 0.188, 0.478}

\newcommand{\ba}{\begin{eqnarray}}
\newcommand{\ea}{\end{eqnarray}}
\def\ov{\over}
\def \del{\partial}

 \def\n{\nu}

\hypersetup{}

\tikzset{Witten diagram/.style={execute at begin picture={%
\draw[blue ,fill=blue!05] circle[radius=\pgfkeysvalueof{/tikz/Witten/radius}];
\path node (X){\phantom{X}};
},baseline={(X.base)}},vertex/.style={circle,fill,inner sep=1.414pt,node
contents={}},
Witten/.cd,radius/.initial=1.414cm}

\begin{document}
\renewcommand{\thefootnote}{\arabic{footnote}}

\overfullrule=0pt
\parskip=2pt
\parindent=12pt
\headheight=0in \headsep=0in \topmargin=0in \oddsidemargin=0in

\vspace{ -3cm} \thispagestyle{empty} \vspace{-1cm}
\begin{flushright} 
\footnotesize
{HU-EP-21/11-RTG}
 
\end{flushright}%

\begin{center}
\vspace{1.2cm}
{\Large\bf \mathversion{bold}
{Mellin amplitudes for $1d$ CFT}\\
}
 
\author{ABC\thanks{XYZ} \and DEF\thanks{UVW} \and GHI\thanks{XYZ}}
 \vspace{0.8cm} {
 Lorenzo~Bianchi$^{a,b}$ \footnote{\tt lorenzo.bianchi@unito.it}, Gabriel~Bliard$^{c}$ \footnote{{\tt gabriel.bliard@physik.hu-berlin.de}},
 Valentina~Forini$^{d,c}$ \footnote{{\tt valentina.forini@city.ac.uk, }}, Giulia~Peveri$^{c}$ \footnote{{\tt giulia.peveri@physik.hu-berlin.de}}}
 \vskip  0.5cm

\small
{\em
$^{a}$  
Dipartimento di Fisica, Universit\`a di Torino and INFN - Sezione di Torino\\ Via P. Giuria 1, 10125 Torino, Italy\\    

$^{b}$  
I.N.F.N. - sezione di Torino,\\
Via P. Giuria 1, I-10125 Torino, Italy\\

$^{c}$  
Institut f\"ur Physik, Humboldt-Universit\"at zu Berlin and IRIS Adlershof, \\Zum Gro\ss en Windkanal 2, 12489 Berlin, Germany\\    

$^{d}$   Department of Mathematics, City, University of London,\\
Northampton Square, EC1V 0HB London, United Kingdom
\vskip 0.02cm
}
\normalsize

\end{center}

\vspace{0.3cm}
\begin{abstract} 
We define a Mellin amplitude for CFT$_1$ four-point functions. Its analytical properties are inferred from physical  requirements on the correlator. We discuss the analytic continuation that is necessary for a fully nonperturbative definition of the Mellin transform. The resulting bounded, meromorphic function of a single complex variable is used to derive an infinite set of nonperturbative sum rules for CFT data of exchanged operators, which we test on known examples. 
We then consider the perturbative setup produced by quartic interactions with an arbitrary number of derivatives in a bulk AdS$_2$ field theory. With our formalism, we obtain a closed-form expression for the Mellin transform of tree-level contact interactions and for the first correction to the scaling dimension of ``two-particle'' operators exchanged in the generalized free field theory correlator. 

\end{abstract}

\newpage

\tableofcontents
 \newpage  
\section{Introduction and discussion}
\label{sec:introduction}

In the one-dimensional case, the conformal invariance of a local field theory is enhanced to an infinite dimensional reparametrization symmetry, resulting in a system with vanishing hamiltonian~\cite{Maldacena:1998uz,Cadoni:1999ja,NavarroSalas:1999up, Sen:2008yk}.  
Relaxing the assumption of locality, i.e. not requiring the presence of a conserved stress tensor, leads to one-dimensional conformal field theories (CFTs) that are invariant under the global conformal group  $SL(2,\mathbb{R})$. The resulting non-local CFTs have a discrete spectrum, making them more similar, in many ways, to their higher dimensional relatives. These theories have explicit realizations in systems such as conformal line defects in higher-dimensional CFTs~\cite{Billo:2013jda, Gaiotto:2013nva, Giombi, defectABJM,Cooke:2017qgm, Kim:2017sju,Bianchi:2018zpb, Giombi:2018qox, Giombi:2018hsx,Kiryu:2018phb, Carlo,  Beccaria:2019dws,Bianchi:2019dlw,Bianchi:2020hsz,  Barrat:2020vch,Ferrero:2021bsb,Agmon:2020pde}, lines of fixed points in SYK models~\cite{Maldacena:2016hyu,Gross:2017vhb},  and in general models defined by the set of boundary correlators of quantum field theories in AdS$_2$~\cite{Paulos:2016fap, Ouyang:2019xdd, Beccaria:2019stp, Beccaria:2019mev,Beccaria:2019dju, Beccaria:2020qtk, Ferrero:2019luz}. These various settings have been recently studied via conformal bootstrap methods, in some cases combined with superspace techniques, or via direct Witten diagrammatics. Furthermore, restricting the kinematics of higher-dimensional CFT correlators leads to a consistent (non-local) $1d$ CFT, so that constraints obtained for one-dimensional systems naturally provide constraints for generic CFTs~\cite{Mazac:2016qev,Mazac:2018mdx,Mazac:2018ycv,Mazac:2018qmi,Kaviraj:2018tfd, Ferrero:2019luz, Paulos:2020zxx}.  

Among all the examples of non-local one-dimensional CFTs, a particularly interesting one for our setting is the defect theory provided by superconformal Wilson lines in $\mathcal{N}=4$ SYM and ABJM theory. In those cases, the set of operator insertions on the line defines the discrete spectrum of a non-local one-dimensional CFT that is believed to be integrable. Indeed, from the seminal work of \cite{Correa:2012hh} and the further developments in \cite{Grabner:2020nis,Cavaglia:2021bnz}, we know that the spectrum of the defect theory can be extracted exactly using fairly standard integrability techniques. Furthermore, on the strong coupling side, we know that this defect theory corresponds to a particular gauge fixing of the classically integrable non-linear sigma model describing the motion of a string in $AdS_5\times S^5$ (or $AdS_4\times \mathbb{CP}^3$ for ABJM). In this context, however, the gauge fixing provides a worldsheet effective field theory in AdS$_2$ background and it is still an interesting open question how the power of integrability can be exploited in this setting. More generally, the study of integrable field theories in curved backgrounds is an active and largely unexplored research subject, which has recently witnessed some interesting developments \cite{Beccaria:2019stp,Beccaria:2019mev,Beccaria:2019dju,Beccaria:2020qtk}. It has been pointed out in many places, see for instance \cite{Giombi, Komatsu:2020sag}, that a crucial ingredient for our understanding of integrability in curved space would be the analogue of flat space S-matrix factorization and we believe Mellin space may provide the correct setting to look for such a feature. The first requirement, however, is a Mellin amplitude which resembles a two-dimensional S-matrix. In this paper, we move the first step in this direction by defining an inherently one-dimensional Mellin transform. 

In the higher-dimensional case, the Mellin representation of conformal correlators~\cite{Mack:2009mi,Penedones:2010ue} has proven to be an excellent tool, especially for the study of holographic CFTs~\cite{Fitzpatrick:2011ia,Paulos:2011ie,Rastelli:2017udc}.
The counting of independent cross-ratios for a $n$-point correlation function  of local operators in a $d$-dimensional CFT  is identical to that of independent variables for a $d+1$-dimensional scattering amplitude. The Mellin representation, or Mellin amplitude, makes this correspondence manifest, expressing the correlators in a form that is the natural AdS counterpart of flat-space scattering amplitudes. 
This construction has several nice features. First, the Mellin amplitude has simple poles located at the values of the twist of exchanged operators (there are, however, infinitely many accumulation points of such poles). Secondly, crossing symmetry of the correlator maps to the amplitude crossing symmetry. Finally, the language of Mellin amplitudes is particularly suitable for large $N$ gauge theories, where perturbation theory is described in terms of Witten diagrams. In this paper we will use these properties as guiding principles for the definition of a Mellin amplitude for $1d$ CFTs.

\setcounter{footnote}{0} 

A four-point correlation function in dimension $d>1$ is a function of two cross-ratios, $(z,\bar z)$, and the associated Mellin amplitude is determined by the usual Mandelstam variables s,t and u,
constrained by the equation $ \text{s+t+u}=\sum_i m_i^2$.
 For $d=1$, instead, the correlator depends on a single cross-ratio, $z$. Correspondingly, two-dimensional $S$-matrices are fully determined in terms of a single Mandelstam variable due to additional kinematical constraints\footnote{
Conservation of energy and momentum for the $2 \to 2 $ scattering process in 2 dimensions leads to the Mandelstam variables
$
\text{s}=(p_1+p_2)^2,\, \text{t}=(p_2-p_3)^2= 4m^2-\text{s},\,\text{u}=(p_3-p_1)^2=0,
$
and the crossing symmetry is written as  $\label{crossingSmatrix}
S(\text{s})=S(4m^2-\text{s})\,$.}. 
One would then naturally expect, for CFT$_1$ correlators, a corresponding Mellin amplitude with the typical features of a two-dimensional $S$-matrix. One option to proceed in the definition of such a Mellin amplitude would be to start from the higher-dimensional definition and enforce the relation $z=\bar z$ (so-called diagonal limit) among cross-ratios. This constraint does not entail a relation in the Mellin variables and thus does not provide an inherently one dimensional Mellin definition. Nevertheless, given a certain Mellin representation of the correlator one can integrate out one of the Mandelstam variables thus obtaining a one-dimensional Mellin transform. A similar approach was followed in~\cite{Ferrero:2019luz}, leading to a successful,  though technically involved, implementation of the Mellin-Polyakov bootstrap~\cite{Polyakov:1974gs,Gopakumar:2016wkt,Gopakumar:2016cpb}. 

In this paper we will  follow a different route and propose a new definition of the Mellin transform, inherently one-dimensional and inspired by the guiding principles outlined above. The general strategy is to infer the analytic properties of the Mellin amplitude $M(s)$ (where $s$ is the complex Mellin variable)  from  physical, nonperturbative requirements on the correlator, which we take to be the one of four identical scalar operators. As we explain in Section \ref{sec:Mellindefinition},  more than one choice is possible and throughout most of our analysis we use the one which displays a transparent correspondence between the dimensions of the operators exchanged in the correlator OPE and the simple poles of the Mellin amplitude. 

Crucially,  the nonperturbative definition requires a finite number of subtractions and analytic continuations, which we perform along the lines of~\cite{Penedones:2019tng}. The Mellin counterpart of the conformal block expansion will provide a clear picture of how to extract the CFT data in Mellin space, while the Regge behaviour of the correlator will impose some powerful bounds on the growth of the Mellin amplitude at large  $s$. Indeed, while the restricted kinematics of the 1$d$ setup does not allow to make contact to several regimes exploited in the higher-dimensional case (e.g. double-light cone limit, Lorentzian OPE limit), one exception is the $u$-channel Regge limit which can be accessed by taking $z,\bar z\to i\infty$ and therefore it is compatible with the diagonal limit $z=\bar z$ \cite{Mazac:2018ycv}. In this limit the correlator should be bounded~\cite{Caron-Huot:2017vep,Maldacena:2015waa,Mazac:2018ycv}, resulting in the boundedness of our Mellin representation at large $s$. 

 
The procedure outlined above and carried out in details in Section~\ref{sec:Mellindefinition} leaves us with a bounded, meromorphic function $M(s)$ in the complex $s$ plane. It satisfies a number of properties,  listed in Section~\ref{derivationsumrules} below, that we can efficiently use to derive a set of nonperturbative sum rules 
for the CFT data of operators that are exchanged in a specific correlation function.
Starting from the pioneering work of \cite{Rattazzi:2008pe},  constraints on the CFT data in the form of sum rules have been derived by the action of specific functionals on the crossing equation. After an impressive numerical effort to exploit them, the one-dimensional analysis of \cite{Mazac:2016qev} provided a first analytical understanding of these functionals, which, after some refinement \cite{Mazac:2018mdx,Mazac:2018ycv,Mazac:2018qmi}, led to a complete set of analytic functionals for $1d$ CFTs. The non-trivial extension to higher dimensions \cite{Mazac:2019shk} paved the way to the beautiful work of \cite{Caron-Huot:2020adz}, where the sum rules obtained using the analytic functionals were shown to be equivalent to those that were derived using the Mellin formalism \cite{Penedones:2019tng,Carmi:2020ekr} or the consistency of the dispersion relation of \cite{Carmi:2019cub}. The distinctive feature of these sum rules, dubbed \emph{dispersive sum rules} in \cite{Caron-Huot:2020adz}, is that they have double zeros at twist $\tau_n=2\Delta_{\phi}+2n$, where $\Delta_{\phi}$ is the dimension of the external scalar operator. Also for $1d$ CFT dispersive sum rules can be derived, using the dispersion relation \cite{Paulos:2020zxx} or using the higher-dimensional definition of the Mellin amplitude \cite{Ferrero:2019luz}. 

In Section~\ref{sumrules} we will start from our definition of Mellin amplitude and derive an infinite set of nonperturbative sum rules. These~\emph{are not} dispersive sum rules, because they have only \emph{single zeros} at $\Delta_n=2\Delta_{\phi}+n$. Such single zeros prevent the presence of positivity properties that are typical of dispersive sum rules.  While such an absence of positivity limits their powerfulness and makes them harder to use within the standard toolkit of the modern conformal bootstrap, we test these sum rules on some known examples and discuss their applicability in a perturbative setting.


The efficiency of the $1d$ Mellin formalism that we propose is manifest in the perturbative setup. Below, in Section~\ref{sec:pert}, we consider first-order deformations from generalized free-field (GFF) theory\footnote{Generalized free-field theory, often called mean field theory, is a non-local theory of operators with generic dimension $\Delta$ and whose correlators are computed by Wick contractions. Since there is no conserved stress tensor, GFF constitutes the simplest possible example of the class of theories we described at the beginning of the Introduction. Its AdS$_2$ dual is the theory of a free massive scalar.} produced by quartic interactions, with any number $L$ of derivatives, in a bulk AdS$_2$ field theory.   We limit our analysis to tree-level contact diagrams, whose building blocks are the D-functions defined in~\cite{Liu:1998ty,DHoker:1999kzh, Dolan:2003hv}. Our strategy is to identify a particularly convenient basis for interactions, in terms of which a correlator corresponding to an arbitrary contact interaction with up to $L$ derivatives is written as a linear combination of reduced $\bar D$-functions of equal weights, see~\eqref{ansatzcorrelator}.  Mellin transforming the associated basis of correlators we are able to obtain the perturbative Mellin amplitude in closed form, equations~\eqref{MellinL0}-\eqref{PDelta}.  This Mellin representation of tree-level contact interactions does not appear to be as simple as in the higher-dimensional case (where it consists of a product of Gamma functions). However, it still allows us to find a closed-form expression for the Mellin transform of the $\bar D$-functions, confirming the  power of the formalism. 

This result can be then efficiently used to extract CFT data. The AdS contact interaction, at this order in perturbation theory, only modifies the CFT data associated to ``two-particle'' operators\footnote{In the literature these operators are also known as double-twist operators or double-trace operators. Since here there is no trace and no twist we opted for the label ``two-particle'' operators. To avoid any confusion, they are unambiguously defined as the conformal primary operators that are exchanged in the GFF correlator. Their scaling dimensions and OPE coefficients are corrected by the perturbation we are considering.} of the symbolic form $\phi\, \square^n \phi$ . This assumption leads, rather non-trivially, to a closed-form expression for the first correction $\gamma^{(1)}_{L,n}$ to their classical dimension in terms of $n$ and the number of derivatives $L$ in the interactions, see equations~(\ref{gammahatsums})-(\ref{gammaLN3}) below. This reproduces existing bootstrap results ($L\leq3$~\cite{Ferrero:2019luz}), and we present a short \texttt{Mathematica} code to easily obtain the explicit expressions for higher values of $L$.  We also present an alternative definition of Mellin transform for which  even simpler results for Mellin amplitudes of contact interactions may be obtained, see Section \ref{sec:alternativeMellin}.  As interesting by-product of this perturbative analysis, some non-trivial identities among D-functions may be obtained - as for example~\eqref{Eq: New relation}. They emerge from noticing that correlators associated to specific derivative interactions must be linear combinations of the ones in the basis that we identified.

The formalism that we propose, once extended to the case of correlators of non-identical operators, finds several natural applications in the context of $1d$ CFTs and AdS$_2$ physics. At the perturbative level, it would be interesting to push our analysis to the case of exchange diagrams and higher-loop corrections.
In particular, given the progress in understanding the analytic structure in cross-ratio space~\cite{Ferrero:2019luz, Ferrero:2021bsb, Alday:2021odx}, it would be interesting to map this knowledge into Mellin space and see whether, as in the higher-dimensional counterpart, implementing the bootstrap directly in Mellin space leads to significant simplifications.

More generally, it is worth analysing the connections of our formalism with the ambitious program of the $S$-matrix bootstrap~\cite{Paulos:2016fap} by carefully studying a proper flat space limit. In this context, it is interesting to notice that, unlike higher-dimensional examples, accumulation points of poles do not seem to appear in this picture. It is therefore important to understand how the singularities of $2d$ $S$-matrices appear in the flat space limit of our Mellin amplitude.

Finally, one important application which strongly motivated our work is the $1d$ CFT defined by supersymmetric Wilson lines in $\mathcal{N}=4$ SYM and ABJM theory. In that case, the dual AdS$_2$ space coincides with the string worldsheet and the effective action is known explicitly. Since the scattering matrix of worldsheet excitations is the standard object which can be computed exactly in the context of AdS/CFT integrability, it would be interesting to analyse how integrability shows up when the worldsheet is AdS$_2$ and whether there is any chance to compute the Mellin amplitude exactly. Notice that, unlike the $S$-matrix for worldsheet excitations around BMN or GKP backgrounds, in this case the Mellin amplitude does not compute the scattering of magnons, but it effectively computes a full four-point function of operators inserted on the Wilson line.
 

\bigskip

The paper proceeds as follows.  After having shortly reviewed kinematical aspects of $1d$ CFTs in Section~\ref{sec:1dcorr}, we present  in Section~\ref{sec:Mellindefinition} our definition of Mellin amplitude discussing its definiteness for nonperturbative correlators in $1d$ CFT. In Section~\ref{sumrules} we derive nonperturbative sum rules for the CFT data of exchanged operators, while in Section~\ref{sec:pert} we discuss the perturbative setting. We conclude with several appendices where we have relegated the technical details.

 \section{Correlation functions in $1d$ CFT}
 \label{sec:1dcorr}

The global conformal group $SL(2,\mathbb{R})$ is generated by translations $P$, dilations $D$ and special conformal transformations $K$ satisfying the commutation relations 
\be
[D,P]=iP, \,\qquad [D,K]=- iK,\,\qquad [K,P]=-2iD.
\ee
The physically relevant irreducible representations are the (unitary) highest-weight representations, corresponding to primary operators labelled by their scaling dimension $\Delta_\phi\geq0$. In $d=1$ there are no rotations and therefore no spin quantum numbers are associated to the primary field\footnote{If the CFT$_1$ is invariant under some internal symmetry group, its local operators are also labeled by the corresponding representations. We will not consider this possibility here.}.

We consider a four-point correlation function of identical scalar operators of dimension $\Delta_{\phi}$. Conformal symmetry imposes\footnote{
While we will consider  fields with bosonic statistics, we notice that in one dimension fermions are just scalar anticommuting fields, and in absence of spin no additional information has to be taken into account beyond their Gra\ss mann nature. In particular, fermionic conformal blocks coincide with the bosonic ones.}
\be\label{corr-id-chi}
\!\!\!\!\!\!
 \braket{\phi(x_1)\phi(x_2)\phi(x_3)\phi(x_4)}=\frac{1}{(x_{12}\,x_{34})^{2\Delta_\phi}} \, \tilde{f}(z)\,,
\ee
where $z$ is the $sl(2, \mathbb{R})$  invariant cross-ratio
\be\label{chi}
 z=\frac{x_{12}\,x_{34}}{x_{13}\,x_{24}} \,,\qquad x_{ij}=x_i-x_j\,\,.
\ee
In $1d$, the order of the operators in the correlation function matters.
Considering the ordering  $x_1\!< x_2 \!< x_3\! < x_4\!\,$, one can use conformal symmetry to fix $x_1=0$, $x_3 = 1$, $x_4 = \infty$, finding that $x_2 \equiv z \in (0, 1)$. Different orderings would generate different ranges for $z$. Unlike the higher dimensional case, the correlators obtained by exchanging $x_1\leftrightarrow x_2$ and $x_1\leftrightarrow x_4$ are not related to \eqref{corr-id-chi} by crossing. For the case of identical operators, one can resort to Bose symmetry and show that the function $\tilde{f}(z)$ still determines the correlators obtained by those exchanges, but these correlators do not coincide with the analytic continuation of $\tilde{f}(z)$ to $z>1$ or to $z<0$. A detailed discussion on this is available in \cite{Mazac:2018qmi}. In this work, we will keep the ordering of the operators fixed, so that we will only be interested in the function $\tilde f(z)$, but we will consider its analytic continuation to complex values. This may seem unphysical thinking of line correlators, but from the perspective of the diagonal limit of higher dimensional correlators it would correspond to Lorentzian regimes for which $\bar z\neq z^*$, but $z=\bar z$. To understand the analytical structure of the function $\tilde f(z)$, we can consider the $s$-channel conformal block expansion
\be\label{OPEs}
\tilde{f}(z)=\sum_{\D} \,c_{\Delta_\phi\Delta_\phi \D}^2 \,\tilde{G}_{\D} (z)\,,\qquad \tilde{G}_{\D}(z)=z^{\D}\, {}_2F_1(\D,\D;2\D;z) \,,
\ee
where  $\D$ is the dimension of the primary operators exchanged in the $\phi\times \phi$ OPE and $c_{\Delta_\phi\Delta_\phi \D}$ are the corresponding OPE coefficients.   $\tilde G_{\D}$  are standard $sl(2)$ blocks resumming the contribution of conformal descendants~\cite{Dolan:2011dv}.
This expansion, accordingly to physical expectations, shows the presence of three branch points at $z=0,1,\infty$. Furthermore, it can be shown that the  OPE \eqref{OPEs}, valid around $z=0$ converges
everywhere but on the branch cuts  $(-\infty, 0]$ and $[1,\infty)$~\cite{Hogervorst:2013sma,Rychkov:2017tpc,Kravchuk:2020scc}. The $t$-channel ($z\to 1$) OPE expansion for $\tilde f(z)$ can be conveniently obtained from the crossing relation
\be \label{cross-z}
\tilde{f}(z)=\left(\frac{z}{1-z}\right)^{2\Delta_\phi}\,\tilde{f}(1-z)\,,
\ee
which is obtained from the symmetry of the correlator under the exchange $x_1\leftrightarrow x_3$, corresponding to $z\to 1-z$~\footnote{Unlike the $x_1\leftrightarrow x_2$ and $x_1\leftrightarrow x_4$ exchanges, the $x_1\leftrightarrow x_3$ is an actual symmetry of the correlator as one can easily see by picturing the four points on a circle. Consistently, this exchange maps the interval $(0,1)$ for $z$ to itself, thus giving a meaning to the relation \eqref{cross-z}}.

In this paper we will conveniently use also an alternative cross-ratio
\be\label{t}
t=\frac{z}{1-z}=\frac{x_{12}\,x_{34}}{x_{14}\,x_{23}}>0\,\,,
\ee
for which crossing maps $t\to 1/t$. The correlator is determined by the function $f(t)$ defined by
\be\label{corr-id-t}
\!\!\!\!\!\!
 \braket{\phi(x_1)\phi(x_2)\phi(x_3)\phi(x_4)}=\frac{1}{(x_{12}\,x_{34})^{2\Delta_\phi}} \, f(t)\,,
\ee
such that $f(t)=\tilde f(\frac{t}{1+t})$. In this case the $s$-channel block expansion for $f(t)$ reads 
\be\label{blockexp}
f(t)= \sum_{\D} \,c_{\Delta_\phi\Delta_\phi \D}^2 \,G_{\D} (t)\,,\qquad G_{\D}(t)=t^{\D}\, {}_2F_1(\D,\D;2\D;-t) 
\ee
and crossing reads 
 \begin{align} \label{cross-f-t}
 f(t)=t^{2\Delta_\phi} \, f\left(\tfrac{1}{t}\right)\,.
\end{align}

For some applications it will be useful to introduce the crossing symmetric function
\be\label{gandf}
\tilde{g}(z)=z^{-2\Delta_\phi} \tilde{f}(z)\, ,\qquad \tilde{g}(z)=\tilde{g}(1-z).
\ee
or, equivalently, in terms of the cross-ratio $t$
\be\label{g_id}
g(t)=\tilde g\left(\tfrac{t}{1+t}\right)\, , \qquad g(t)=  g(\tfrac{1}{t})\,.
\ee

There is another interesting limit we will consider in the following, i.e. the~$z\to \frac12+ i \infty$ limit (we could take this limit along any direction excluding the real line to avoid the branch cuts, but for definiteness we take it along the imaginary axis). This limit can be understood in terms of the higher-dimensional correlator in the diagonal limit, where it corresponds to the $u$-channel Regge limit\footnote{In $1d$ there is no $u$-channel OPE expansion as it is impossible to bring $x_1$ close to $x_3$ without $x_2$ in between. However, one can resort on the higher dimensional picture to understand that while the $u$-channel OPE would correspond to $z\to i\infty$ and $\bar z\to -i\infty$, the $u$-channel Regge limit is $z,\bar z\to i\infty$.}. In particular,  four-point functions of a unitary CFT are bounded in the Regge limit \cite{Caron-Huot:2017vep,Maldacena:2015waa} and we have \cite{Mazac:2018ycv}
\be\label{Reggebound}
\left|\tilde{g}\left(\textstyle{\frac{1}{2}}+iT\right)\right|\, \text{is bounded as } T\rightarrow \infty.
\ee
Translating into the $t$ cross-ratio \eqref{t}, the line parametrized by $z=\frac12+i \,\xi$ is mapped into the unit circle $t=e^{i \theta}$ for $\theta\in(-\pi,\pi)$ and the Regge limit occurs when $\theta\to \pi$. The Regge boundedness condition \eqref{Reggebound} for the function $f(t)$ in~\eqref{corr-id-t} then reads
\begin{equation}\label{reggetheta}
 f(e^{i\theta})=\mathcal{O}\left((\pi-\theta)^{-2\Delta_\phi}\right) \qquad \theta\to \pi .
\end{equation}
Further details on the implication of the Regge boundedness condition on the Mellin amplitude can be found in Section \ref{boundedness}.

 \section{A Mellin transform for correlators in $1d$ CFT} 
 \label{sec:Mellindefinition}
  
Since the $1d$ correlator depends on a single cross-ratio, it is useful to start by looking at the textbook Mellin transform for a function $F(t)$ defined on the positive real axis
\begin{align}\label{textbook}
 \mathcal{M}[F](s)=\int_0^{\infty} \! d t\,  F(t)\, t^{-1-s} \,  .
\end{align}
Given this definition, it is natural to consider  for $t$ the cross-ratio  in~\eqref{t}, which is defined in the correct range. Furthermore, one has to choose which function of the cross-ratio should be identified with $F$ in the definition \eqref{textbook}. As we described in Section~\ref{sec:1dcorr} different choices of the prefactor in \eqref{corr-id-chi} lead to different functions of the cross-ratio, related to each other by powers of $t$ and $(1+t)$. In contrast to the higher dimensional case, where a rescaling by powers of the cross-ratios has the effect of shifting the corresponding Mellin variables, in one dimension, a rescaling by powers of $t$ leads to a shift in $s$, whereas a rescaling by powers of $(1+t)$ leads to different Mellin amplitudes. This presents us with the question of which criterium one should use to define the Mellin amplitude. Up to shifts in the $s$ variable, we define a one-parameter family of Mellin amplitudes given by
\begin{align}\label{Mellin1param}
 \mathcal{M}_a(s)=\int_0^{\infty}\!\! dt \, f(t) \Big(\frac{t}{1+t}\Big)^a\,t^{-1-s} \, ,
\end{align}
where the function $f(t)$ is given in \eqref{corr-id-t}. Using the crossing relation \eqref{cross-f-t}, one immediately finds the functional relation for the Mellin amplitude
\be\label{crossingmellin}
\mathcal{M}_a(s)=\mathcal{M}_a(2\Delta_\phi+a-s)\,.
\ee
Clearly, \eqref{crossingmellin} is reminiscent of the crossing for $S$-matrix elements in two dimensions, see footnote~\ref{crossingSmatrix}. However, the precise relation between $s$ and the ordinary flat space Mandelstam variable~s requires a careful analysis of the flat space limit, which we do not address in this work\footnote{Here we just notice that the large $s$ regime is the relevant one for the flat space limit considered in~\cite{Penedones:2010ue}, where AdS scattering reduces to the scattering of massless excitations for large AdS radius. In that case, one would have the flat space relation $\mathcal{M}_a(s)=\mathcal{M}_a(-s)$ for any finite value of $a$. This relation would be consistent with 2d massless scattering, where $\text{s}=-\text{t}$. There is however more than one approach to the flat space limit, see~\cite{Paulos:2016fap}.}. Up to shifts in the $s$ variable, the definition \eqref{Mellin1param} allows for different choices of prefactors in the correlator \eqref{corr-id-t}. For instance, the choice $a=0$ clearly corresponds to taking the Mellin transform of $f(t)$, while the choice $a=-2\Delta_\phi$ effectively corresponds to taking the Mellin transform of the function $g(t)$ in \eqref{g_id}. In the following, we will mostly focus on the choice $a=0$, which emerges naturally when considering the $s$-channel conformal block expansion.
In Section~\ref{sec:alternativeMellin}, we introduce the possible alternative $a=-2\Delta_\phi+1$ which leads to simple results in a perturbative expansion around GFF.

Let us briefly comment on the relation between the definition \eqref{Mellin1param} and the diagonal limit of the higher-dimensional Mellin transform. In higher dimensions, the Mellin transform involves a double integration over the two cross-ratios $u$ and $v$ for $0<u,v<\infty$. One could naively impose by hand the diagonal limit condition $\sqrt{u}+\sqrt{v}=1$ on the integral and correspondingly identify the Mellin variables with a relation of the kind $\mathrm{s}+\mathrm{t}=4\Delta_{\phi}$. The form of the integrand would then fall in the class described by \eqref{Mellin1param}, but the range of the integration would include negative values of $t$ (this is because the diagonal limit condition identifies a curve in the $u,v$ plane which is parametrized by real values of $t$). As we discussed in Section \ref{sec:1dcorr}, the kinematics of one-dimensional correlators is subtle and the correlator is well defined only for a specific region of the cross ratio \footnote{For the case of identical operators one can resort to Bose symmetry and give a meaning to the correlator on the whole $t$-real axis, but since we are after a more general definition we did not follow this route here.}. We then decided to restrict the integration contour to the part of the real axis where the correlator is well defined.

\subsection{Nonperturbative Mellin amplitude} \label{sec:nonpert}

We start with the Mellin amplitude $\mathcal{M}_0(s)$ defined in \eqref{Mellin1param}, which we multiply by an overall factor for future convenience
\be\label{Mellin}
M(s) =\frac{1}{\Gamma(s)\Gamma(2\Delta_\phi-s)}\int_0^\infty dt\, f(t)\, t^{-1-s}\,.
\ee 
In this case the crossing relation \eqref{crossingmellin} reads
\be\label{crossing1def}
M(s)=M(2\Delta_\phi -s)\, .
\ee
The goal of our discussion is to infer the analytic properties of the Mellin amplitude $M(s)$ from the physical requirements on the correlator $f(t)$. First, following \cite{Penedones:2019tng}, we remind a general result for the one-dimensional Mellin transform \eqref{Mellin}.

\subsubsection{A theorem}\label{theorem}
Consider a function $F(t)$ in the vector space $\mathcal{F}_H^\Theta$ of complex valued functions that are holomorphic for $\textrm{arg}(t)\in \Theta$ and obey
\begin{equation}\label{boundnessF}
	|F(t)| \leq  \frac{C(h)}{|t|^h} \qquad h \in H \, ,
\end{equation}
where H is a subset of $\mathbb{R}$, typically of the form $H=(h_{min}, h_{max})$. Consider also the function $\hat{M}(s)$ in the vector space $\mathcal{M}_{H}^{\Theta}$ of complex valued functions that are holomorphic for $\text{Re}(s)\in H$ and exponentially suppressed in the limit $|\text{Im}(s)|\rightarrow  \infty$
\begin{align} \label{bound M}
	|\hat{M}(s)| \leq K(\textrm{Re}(s))e^{-|\text{Im}(s) \sup_{\Theta} \textrm{arg}(t)|} \qquad |\text{Im}(s)|\to \infty \, .
\end{align}
These two vector spaces exists independently, but the following theorem holds
\begin{theorem}
 Given a function $F(t)\in \mathcal{F}_H^\Theta$, its Mellin transform $\mathcal{M}[F](s)$ exists and $\mathcal{M}[F](s)\in \mathcal{M}_{H}^{\Theta}$. Furthermore, $\mathcal{M}^{-1}\mathcal{M}[F](t)=F(t)$ for any $\arg (t)\in \Theta$. Conversely, given $\hat{M}(s)\in \mathcal{M}_{H}^{\Theta}$ its inverse Mellin transform exists and $\mathcal{M}^{-1}[\hat{M}](s)\in \mathcal{F}_H^\Theta$. Furthermore, $\mathcal{M}\mathcal{M}^{-1}[\hat{M}](s)=\hat M(s)$ for any $s\in H+i\mathbb{R}$.
\end{theorem}

This is a classical result for the one-dimensional Mellin transform so we are not going to prove it here. Instead, we will discuss how the physical $1d$ correlator violates the hypothesis of the theorem and how we can overcome this issue. The convergence of the $s$-channel OPE for $|\arg(t)|<\pi$ ensures that the function $f(t)$ is indeed analytic in a sectorial domain $\Theta$. Nevertheless, the condition \eqref{boundnessF} is violated in two ways
\begin{itemize}
 \item When light operators ($\Delta<\Delta_\phi$) are exchanged in the OPE, the region $H$ is not well defined and the Mellin transform does not exist. This issue is analogous to the higher dimensional case of \cite{Penedones:2019tng} and we will solve it by implementing a finite number of subtractions in Section \ref{convandsub}.
 \item The correlator $f(t)$ is not bounded for $t\to e^{i \pi}$ where it has a singularity controlled by the Regge limit \eqref{reggetheta}. This issue does not spoil the existence of the Mellin transform, but it gives a result that is not bounded by \eqref{bound M}. 
\end{itemize}
To understand this second point let us present a simple example which will be useful to explain the issue. Consider the function $F(t)=(\frac{t}{1+t})^{2\Delta_{\phi}}$. This function is analytic for $|\arg(t)|<\pi$ and gives a convergent integral \eqref{Mellin} for $0<\text{Re}(s)<2\Delta_{\phi}$. However, despite the bound \eqref{boundnessF} holds along the real axis, it is violated for $t\to e^{i \pi}$. The Mellin transform of this function reads
\begin{equation}\label{example}
 \int_{0}^{\infty} dt \left(\frac{t}{1+t}\right)^{2\Delta_{\phi}} t^{-1-s}=\frac{\Gamma(s)\Gamma(2\Delta_{\phi}-s)}{\Gamma(2\Delta_{\phi})} \,  .
\end{equation}
From this explicit expression we see immediately that for $|\text{Im}(s)|\to \infty$ the r.h.s.  is not bounded by $e^{-\pi |\text{Im}(s)|}$. Rather, it is bounded by
\begin{equation}
 \frac{\Gamma(s)\Gamma(2\Delta_{\phi}-s)}{\Gamma(2\Delta_{\phi})}\leq K(\textrm{Re}(s))  |\text{Im}(s)|^{2\Delta_{\phi}-1} e^{-\pi |\text{Im}(s)|} \qquad |\text{Im}(s)|\to \infty \, .
\end{equation}
As we see, the exponential decay is correctly predicted by the theorem, while the additional polynomial divergence can be related to the behaviour of the function $F(t)$ for $t\to e^{i\pi}$. 
In Section \ref{boundedness}, we will show that this is a specific instance of a general relation between the large $s$ asymptotics of $M(s)$ to the Regge limit of $f(t)$.

\subsubsection{Convergence and subtractions}\label{convandsub}
Let us now discuss the convergence of the integral \eqref{Mellin}. Let $f(t)$ be well behaved for $t\in \mathbb{R}^+$, that is, we do not want divergences in $t$ other than at $t=0$ and $t\rightarrow \infty$. This behaviour coincides with that of the CFT$_1$ correlators we will be interested in. Consider the behaviour of $f(t)$  close to $t=0$. Using~\eqref{blockexp}, we find that the leading power is $f(t)\sim t^{\D_0}$ with $\Delta_0$ the dimension of the lightest exchanged operator. 
Analogously, using the crossing symmetry relation~\eqref{cross-f-t}, we find that the large $t$ behaviour of $f(t)$ is $f(t)\sim t^{2\Delta_\phi-\D_0}$. 
Therefore the integral converges in the strip
\be\label{convergence}
2\Delta_\phi-\D_0<\text{Re}(s)<\D_0\,, 
\ee
which is a well-defined interval \emph{only for $\D_0>\Delta_\phi$}.  
In order to give a nonperturbative definition of the Mellin transform, which allows for lighter operators to be exchanged,  we need to perform some subtractions along the lines of~\cite{Penedones:2019tng}\footnote{See in particular Appendix B in~\cite{Penedones:2019tng} for the one-dimensional case.}. One obvious example is GFF, where the identity operator is exchanged. We will consider this case explicitly in Section~\ref{sec:GFF}. For the moment, we consider the Mellin transform of the connected part of the correlator. Let us consider the following subtractions
\begin{align} \label{subtr0}
 f_0(t)&=f_{\text{conn}}(t)-\sum_{\D_{0}\leq \D \leq\Delta_\phi}\sum_{k=0}^{[\Delta_\phi-\D]} c_\D \frac{(-1)^{k}}{k!} \frac{\Gamma(\D+k)^2\Gamma(2\D)}{\Gamma(\D)^2\Gamma(2\D+k)} t^{\D+k}\,,\\
 f_{\infty}(t)&=f_{\text{conn}}(t)-\sum_{\D_{0}\leq \D \leq\Delta_\phi}\sum_{k=0}^{[\D_\phi-\D]} c_\D\frac{(-1)^k}{k!} \frac{\Gamma(\D+k)^2\Gamma(2\D)}{\Gamma(\D)^2\Gamma(2\D+k)} t^{2\Delta_\phi-\D-k}\,,\label{subtrinf}
\end{align}
where, for convenience, we write  $c_{\Delta_\phi \Delta_\phi \Delta}^2\equiv c_{\Delta}$.
For the function $f_0(t)$ we subtracted the $s$-channel contribution of all the operators (primaries and descendants) with scaling dimension below the threshold $\D=\Delta_\phi$, making use of the series expansion of the hypergeometric function in~\eqref{blockexp}. This improves the behaviour of the function at $t=0$. On the other hand, for $f_{\infty}(t)$ we subtracted all the $t$-channel operators below threshold, thus improving the behaviour at $t=\infty$. The idea is to split the integral~\eqref{Mellin} in two parts, which are defined on (possibly non-overlapping) semi-infinite regions of the complex $s$ plane
\begin{align}\label{psi0def}
 \psi_0(s)&=\int_0^1 dt\, f_{\text{conn}}(t) \,t^{-1-s} & \text{Re}(s)&< \D_0 \,  , \\
  \psi_{\infty}(s)&=\int_1^{\infty} dt \,f_{\text{conn}}(t) \,t^{-1-s} & \text{Re}(s)&>2\Delta_\phi- \D_0 \label{psiinfdef}
\end{align}
When the two regions do not overlap, we analytically continue $\psi_0(s)$ and $\psi_{\infty}(s)$ by considering the integrals of the functions  \eqref{subtr0} and \eqref{subtrinf} and adding a finite number of poles
\begin{align}
\label{psi0ancont}
\!\!\!\! \psi_0(s)&=\int_0^1 \!\!\!dt \, f_0(t)\, t^{-1-s}+\!\!\!\!\!\!\sum_{\D_{0}\leq \D \leq\Delta_\phi}\!\!\! \! \sum_{k=0}^{[\Delta_\phi-\D]} c_\D \tfrac{(-1)^{k}}{k!} \tfrac{\Gamma(\D+k)^2\Gamma(2\D)}{\Gamma(\D)^2\Gamma(2\D+k)} \tfrac{1}{s-\D-k}\,, \qquad \small{\text{Re}(s)< \tilde \D_0} \, , \\
\!\!\!\! \psi_{\infty}(s)&=\int_1^{\infty} \!\!\!\!\!\!dt \, f_{\infty}(t)\, t^{-1-s}+\!\!\!\!\!\!\sum_{\D_{0}\leq \D \leq\Delta_\phi}\!\!\! \!\sum_{k=0}^{[\Delta_\phi-\D]} c_\D \tfrac{(-1)^{k}}{k!} \tfrac{\Gamma(\D+k)^2\Gamma(2\D)}{\Gamma(\D)^2\Gamma(2\D+k)}  \tfrac{1}{s-2\Delta_\phi+\D+k}\,,\small{\text{Re}(s)>2\Delta_\phi- \tilde \D_0} \, ,\label{psiinfancont}
\end{align}
where $\tilde \D_0>\Delta_\phi$ is the lightest exchanged operator above threshold (notice that this operator could be either a primary or a descendant). Both these functions are now well defined on the non-vanishing strip $2\Delta_\phi- \tilde \D_0<\text{Re(s)}<\tilde \D_0$ and therefore their sum yields a well defined Mellin transform
\begin{align}\label{Mstrip}
 M(s)&=\frac{\psi_0(s)+\psi_{\infty}(s)}{\Gamma(s)\Gamma(2\Delta_\phi-s)} \, ,& 2\Delta_\phi- \tilde \D_0&<\text{Re(s)}<\tilde \D_0 \, .
\end{align}
The price to pay is a deformation of the integration contour in the inverse Mellin transform, which reads
\begin{align}\label{inverseMellin}
f(t) = \int_\mathcal{C}\frac{ds}{2 \pi i}\,\Gamma(s)\Gamma(2\Delta_\phi-s)\,M(s)\,t^{s}\, .
\end{align}
To understand the form of the contour $\mathcal{C}$, we need to discuss the analytic structure of $M(s)$. In order to do this, one can follow the strategy described above to extend the definition \eqref{Mstrip} to the whole complex $s$ plane. To analytically continue $\psi_0(s)$ from the region $\text{Re}(s)<\D_0$ to the region $\text{Re}(s)<\tilde \D_0$ we subtracted a few exchanged operators in $f(t)$ and added a finite number of poles in \eqref{psi0ancont}. By adding more and more poles, we can further extend the area of analyticity. We then conclude that the \emph{Mellin block expansion} defined by
\begin{equation}\label{Mellinpoles} 
 M(s)=\frac{\psi_0(s)+\psi_{\infty}(s)}{\Gamma(s)\Gamma(2\Delta_\phi-s)}
\end{equation}
with
\begin{align}\label{psi0sum}
 \psi_0(s)&=\sum_{\D}\sum_{k=0}^{\infty} c_\D \frac{(-1)^{k+1}\Gamma(\D+k)^2\Gamma(2\D)}{k!\Gamma(\D)^2\Gamma(2\D+k)}\frac{1}{s-\D-k} \, ,\\
 \psi_{\infty}(s)&=\sum_{\D}\sum_{k=0}^{\infty} c_\D \frac{(-1)^{k}\Gamma(\D+k)^2\Gamma(2\D)}{k!\Gamma(\D)^2\Gamma(2\Delta+k)}\frac{1}{s-2\Delta_\phi+\D+k} \label{psiinfsum}
\end{align}
provides a representation of $M(s)$ which is valid on the \emph{whole complex $s$ plane} (excluding the point at infinity which will be discussed in details in Section \ref{boundedness}). In particular, the representation~\eqref{Mellinpoles} immediately allows us to read off the position of the poles of $M(s)$\footnote{In principle there could be an additional singularity at $\infty$, but we postpone this discussion to Section~\ref{derivationsumrules}.}. For any exchanged primary operator of dimension $\D$ there are two infinite sequence of poles running to the right of $s=\D$ and to the left of $s=2\Delta_\phi-\D$. Following the common nomenclature we denote them as
\begin{align}
\text{\emph{right} poles}: s_R&=\D+k\,, ~~\qquad\,\, \qquad k=0,1,2,\dots \label{leftpoles}\\
\text{\emph{left} poles}: s_L&=2\Delta_\phi-\D-k\,, \qquad k=0,1,2,\dots \label{rightpoles}\\
\text{Res}[M(s)]|_{s_L}&\equiv-\text{Res}[M(s)]|_{s_R}=\frac{(-1)^{k} \Gamma (2 \Delta) \Gamma (\D+k)}{k! \,\Gamma (\D)^2 \Gamma (2 \D+k) \Gamma (2   \Delta_\phi-\D-k )}\,. \label{residues}
\end{align}
Notice that the precise identification of the sum over $k$ in \eqref{psi0sum} with the sum over descendants in the block expansion is a consequence of the choice $a=0$ in \eqref{Mellin1param}. Different choices of $a$ in \eqref{Mellin1param} would lead to a less transparent correspondence between poles and conformal descendants.

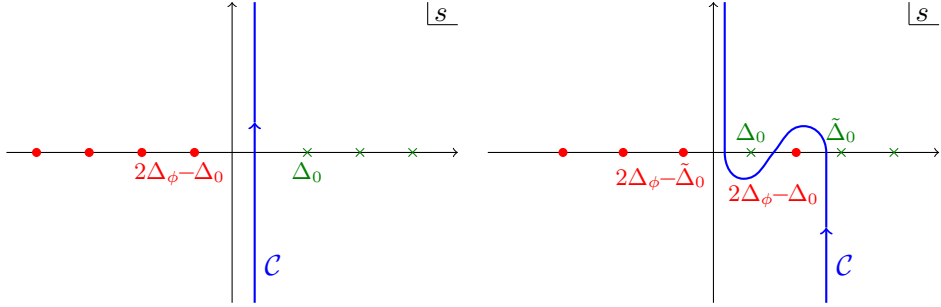
\begin{figure} 
		\center
		\begin{tikzpicture}
			\draw[->](-3,0)--(3,0);
			\draw[->](0,-2)--(0,2);
			\draw[] (2.6,2)--(2.6,1.7);
			\draw[] (3,1.7)--(2.6,1.7);
			\node[anchor = north east] at (3,2.05) {$s$};
			\draw[thick,->,blue] (0.3,-2)--(0.3,0.4);
			\draw[thick,blue] (0.3,0.4)-- (0.3,2);
			\node[blue, right] at (0.3,-1.5) {$\mathcal{C}$};
			\filldraw [red] (-0.5,0) circle (1.5pt);
			\filldraw [red] (-1.2,0) circle (1.5pt);
			\filldraw [red] (-1.9,0) circle (1.5pt);
			\filldraw [red] (-2.6,0) circle (1.5pt);
			\draw (1,0) node[cross=2pt,green!50!black] {};
			\draw (1.7,0) node[cross=2pt,green!50!black] {};
			\draw (2.4,0) node[cross=2pt,green!50!black] {};
			\node[below, green!50!black] at (1,0) {\footnotesize{$\D_0$}};
			\node[below, red] at (-0.7,0) {\footnotesize{$2\Delta_\phi\!\!-\!\! \D_0$}};
		\end{tikzpicture}\quad
		\begin{tikzpicture}
			\draw[->](-3,0)--(3,0);
			\draw[->](0,-2)--(0,2);
			\draw[] (2.6,2)--(2.6,1.7);
			\draw[] (3,1.7)--(2.6,1.7);
			\node[anchor = north east] at (3,2.05) {$s$};
			\node[blue, right] at (1.5,-1.5) {$\mathcal{C}$};
			\filldraw [red] (1.1,0) circle (1.5pt);
			\filldraw [red] (-0.4,0) circle (1.5pt);
			\filldraw [red] (-1.2,0) circle (1.5pt);
			\filldraw [red] (-2,0) circle (1.5pt);
			\draw (0.5,0) node[cross=2pt,green!50!black] {};
			\draw (1.7,0) node[cross=2pt,green!50!black] {};
			\draw (2.4,0) node[cross=2pt,green!50!black] {};
			\node[above, green!50!black] at (0.5,0) {\footnotesize{$\D_0$}};
			\node[below, red] at (.8,-0.3) {\footnotesize{$2\Delta_\phi\!\!-\!\! \D_0$}};
            \node[above, green!50!black] at (1.7,0) {\footnotesize{$\tilde \D_0$}};
			\node[below, red] at (-0.7,0) {\footnotesize{$2\Delta_\phi\!\!-\!\! \tilde \D_0$}};
			\draw [->,thick,blue] (1.5,-2) to[out=90,in=-90](1.5,-1) ;
			\draw [,thick,blue] (1.5,-1) to[out=90,in=-90](1.5,0) to[out=90,in=0](1.2,.35) to [out=180, in=50] (0.8,0)to [out=50-180, in =0](.4,-.35) to[out=180, in=-90] (.15,0)to[out=90, in = -90](.15,2);
		\end{tikzpicture}\quad
		\caption{\textbf{Left}: The contour for the inverse Mellin transform when $\D_0>\Delta_\phi$. Left poles are marked in \textcolor{red}{red} and right poles are \textcolor{green!50!black}{green}. \textbf{Right}: When $\D_0<\Delta_\phi$ left and right poles intersect and the contour needs to be deformed.}\label{fig:C}
	\end{figure}

Given this structure of poles, we can now give a precise definition of the contour $\mathcal{C}$ in \eqref{inverseMellin}. The contour $\mathcal{C}$ is chosen in such a way to leave all the \emph{right} poles of $M(s)$ on its right and all the \emph{left} poles on its left. If the lightest exchanged operator has dimension $\D_0>\Delta_\phi$, no analytic continuation is required in \eqref{psi0def} and \eqref{psiinfdef} (in other words the set of left and right poles do not overlap) and any contour within the interval \eqref{convergence} will suffice, see for example the straight one on the left in Fig.\ref{fig:C}. 

When lighter operators are exchanged, the contour needs to be deformed because the set of right poles intersects with the set of left poles. In Figure~\ref{fig:C} we show an example with a single operator below threshold. It is clear from the picture that a more complicated situation arises when a left and a right pole coincide. This happens, for instance, for the GFF case, which we address in Section~\ref{sec:GFF}. More generally, this happens whenever there is an exchanged operator with dimension $\D=\Delta_\phi+\frac{\mathbb{Z}}{2}$. In a generic spectrum we do not expect this to be the case.

We conclude this section by noticing that we can perform the sum over $k$ in~\eqref{Mellinpoles}, resumming all the conformal descendants in a crossing symmetric Mellin block expansion 
\begin{align} \label{hypergeom}
 M(s) &= \frac{1}{\Gamma(s)\Gamma(2\Delta_\phi-s)} \sum_\D\,c_\D [\mathcal{F}_\D(s)+\mathcal{F}_\D(2\Delta_\phi-s)]\,,\\
\mathcal{F}_\D(s)&= \frac{{}_3F_2({\D, \D, \D - s}; {2 \D,1+ \D-s}; -1)}{\D-s} \, .
\end{align}
Before analysing the behaviour of $M(s)$ for $s\to \infty$, we consider a simple yet subtle example, that of generalized free field theory.

\subsubsection{A degenerate example: generalized free field theory}
\label{sec:GFF}
\begin{figure}
		\center
		\begin{tikzpicture}
			\draw[->](-3,0)--(3,0);
			\draw[->](0,-2)--(0,2);
			\draw[] (2.6,2)--(2.6,1.7);
			\draw[] (3,1.7)--(2.6,1.7);
			\node[anchor = north east] at (3,2.05) {$s$};
			\filldraw [red] (1,0) circle (1.5pt);
            \filldraw [red] (0,0) circle (1.5pt);
			\filldraw [red] (-0.7,0) circle (1.5pt);
			\filldraw [red] (-1.4,0) circle (1.5pt);
			\filldraw [red] (-2.1,0) circle (1.5pt);
			\draw (0,0) node[cross=2pt,green!50!black] {};
			\draw (1,0) node[cross=2pt,green!50!black] {};
			\draw (1.7,0) node[cross=2pt,green!50!black] {};
			\draw (2.4,0) node[cross=2pt,green!50!black] {};
			\node[below] at (1,0) {\footnotesize{$2\Delta_\phi$}};
		\end{tikzpicture}\quad
		\begin{tikzpicture}
			\draw[->](-3,0)--(3,0);
			\draw[->](0,-2)--(0,2);
			\draw[] (2.6,2)--(2.6,1.7);
			\draw[] (3,1.7)--(2.6,1.7);
			\node[anchor = north east] at (3,2.05) {$s$};
			\node[blue, right] at (1.1,-1.5) {$\mathcal{C}$};
			\filldraw [red] (1,0) circle (1.5pt);
            \filldraw [red] (0,0) circle (1.5pt);
			\filldraw [red] (-0.7,0) circle (1.5pt);
			\filldraw [red] (-1.4,0) circle (1.5pt);
			\filldraw [red] (-2.1,0) circle (1.5pt);
			\draw (0.2,0) node[cross=2pt,green!50!black] {};
			\draw (1.2,0) node[cross=2pt,green!50!black] {};
			\draw (1.9,0) node[cross=2pt,green!50!black] {};
			\draw (2.6,0) node[cross=2pt,green!50!black] {};
			\node[above, green!50!black] at (0.3,0) {\footnotesize{$\e$}};
			\node[below, red] at (0.8,0) {\footnotesize{$2\Delta_\phi$}};
            \node[above, green!50!black] at (1.5,0) {\footnotesize{$2\Delta_\phi\!\!+\!\!\e$}};
			\draw [->,thick,blue] (1.1,-2) to[out=90,in=-90](1.1,-1) ;
			\draw [,thick,blue] (1.1,-1) to[out=90,in=-90](1.1,0) to[out=90,in=0](.9,.35) to [out=180, in=60] (0.6,0)to [out=60-180, in =0](.3,-.35) to[out=180, in=-90] (.1,0)to[out=90, in = -90](.1,2);
		\end{tikzpicture}\quad
		\caption{\textbf{Left}: The configuration of the poles for GFF. The left and right poles coincide in $s=0$ and $s=2 \Delta_\phi$. \textbf{Right}: the right poles are slightly shifted to the right so that the contour $\mathcal{C}$ can run in the middle.}\label{fig:GFFpoles}
	\end{figure}
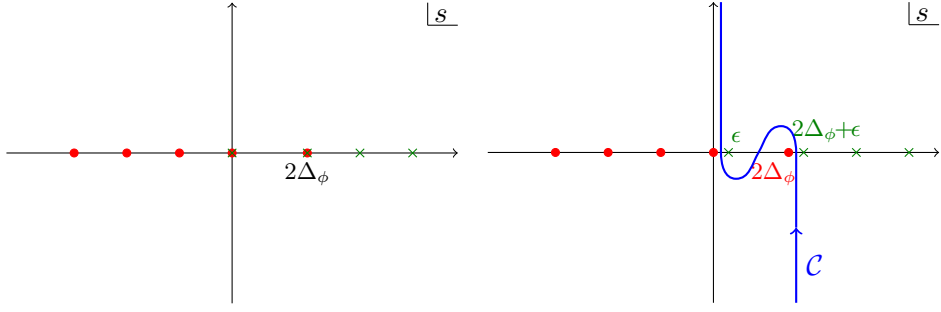

Let us consider the simplest possible example of $1d$ CFT. The GFF correlator for four identical scalars of dimension $\Delta_\phi$ reads
\begin{equation}\label{fGFF}
 f^{\text{GFF}}(t)=1+ t^{2\Delta_\phi}+\left(\frac{t}{1+t}\right)^{2\Delta_\phi}.
\end{equation}
Of course in this case we know very well the spectrum of exchanged operators, which includes the identity,  ``two-particle'' operators of the schematic form 
\be\label{doubletrace}
[\phi\phi]_n\sim\phi\, \square^n \phi,
\ee 
and their conformal descendants. The conformal primary operators have dimension $\D_n=2\Delta_\phi+2n$ for $n\geq 0$. As usual, the fact that the dimensions of these operators are separated by integers creates huge degeneracies. In particular, we should be worried by the coincidence of left and right poles in \eqref{leftpoles} and \eqref{rightpoles}. This happens for $s=0$, where the first right pole, associated to the exchange of the identity in the $s$-channel, coincides with the second left pole, associated to the exchange of $\phi^2$ in the $t$-channel, and at $s=2\Delta_\phi$ for the crossing symmetric case. The situation is depicted in Figure \ref{fig:GFFpoles}.

Coincident poles generate a problem in the realization of the contour $\mathcal{C}$, which was defined precisely in such a way to separate left and right poles. To avoid this issue one can slightly split the poles by shifting the right (or left) poles by a small amount, as shown in Figure \ref{fig:GFFpoles}. The result of the procedure outlined in Section \ref{sec:nonpert} leads to a Mellin transform which depends on the shift $\e$
\begin{equation}\label{resultGFFepsilon}
 M^{\text{GFF}}_{\e}(s)=\frac{1}{\Gamma(s)\Gamma(2\Delta_\phi-s)}\left(\frac{\Gamma(s)\Gamma(2\Delta_\phi+\e-s)}{\Gamma(2\Delta_\phi)}+\frac{1}{s-\e}+\frac{1}{s-2\Delta_\phi-\e}-\frac{1}{s}-\frac{1}{s-2\Delta_\phi}\right) \, .
\end{equation}
Taking the inverse Mellin transform \eqref{inverseMellin} with the contour $\mathcal{C}$ depicted in Figure \ref{fig:GFFpoles} and then taking the limit $\e\to0$, one recovers \eqref{fGFF}. Taking the limit $\e\to0$ before integrating leads to the very simple result $M_0(s)=\frac{1}{\Gamma(2\Delta_\phi)}$, however the contour $\mathcal{C}$ is not well defined and one cannot consistently take the inverse Mellin transform\footnote{However, the correction to this is of order $M_{\epsilon}(s)-M_0(s) =\frac{\epsilon}{s^2}+\frac{\epsilon}{(s-2\Delta_\phi)^2}$ which is only non-negligible in the $s\rightarrow 0$ and $s\rightarrow2\Delta_\phi$ limit. Therefore, for all other purposes (for example for comparisons at perturbative level, or sum rule results), we can effectively take the $M^{GFF}_0(s)$ result.}. Notice that this procedure is simply the limit in the weak sense that is normally considered in distribution theory. In particular, the identity 
\begin{equation}
\lim_{\e\to0}\int_{p-i \infty}^{p+i\infty}\frac{ds}{2\pi i } \left(\frac{1}{s-p-\e}-\frac{1}{s-p+\e}\right) t^{s}=t^p
\end{equation}
tells us that the function $\frac{1}{s-p-\e}-\frac{1}{s-p+\e}$ in the limit $\e\to 0$ defines a delta function $\delta(s-p)$ for Mellin integration\footnote{Attentive readers will probably have noticed that the individual terms in $\lim_{\e\to0}\int_{p-i \infty}^{p+i\infty}\frac{ds}{2\pi i } (\frac{1}{s-p-\e})$ diverge and one cannot use the residue theorem . However, the combination does not have this problem since $\left(\frac{1}{s-p-\e}-\frac{1}{s-p+\e}\right) = \frac{2\epsilon}{(s-p)^2-\e^2}$ which insures convergence at large s.}. In this sense, we can say that the Mellin transform for GFF is a distribution rather than a function.
This issue is clearly caused by the exchange of the identity operator and by the degeneracy typical of GFF. 
An alternative route for obtaining the Mellin transform of the GFF correlator would be to use equations \eqref{psi0sum} and \eqref{psiinfsum} with the well-known expressions of the GFF OPE coefficients
\begin{equation}\label{GFFOPE}
 c^{(0)}_n =\frac{2 \Gamma (2 n+2 \Delta_\phi )^2 \Gamma (2 n+4 \Delta_\phi -1)}{\Gamma (2 \Delta_\phi )^2 \Gamma (2 n+1) \Gamma (4 n+4 \Delta_\phi -1)}
\end{equation}
and summing over $\D_n=2\Delta_\phi+2n$. Doing this one finds again the result $M(s)=\frac{1}{\Gamma(2\Delta_\phi)}$ consistently with the fact that equations \eqref{psi0sum} and \eqref{psiinfsum} were obtained under the assumption of non-coincident poles. Also in this case, one could slightly shift the position of the poles in $\psi_{0}(s)$ (or $\psi_{\infty}(s)$) and recover an expression which coincides with $M_{\e}^{GFF}$ in the weak limit $\e\to 0$. 


\subsection{Regge limit and Mellin boundedness} \label{boundedness}
In this section we will derive a bound on the large $s$ behaviour of the Mellin amplitude $M(s)$ using the Regge behaviour of the function $f(t)$, i.e. the limit $t\to e^{i\pi}$ described in \eqref{reggetheta}. 
Looking at the direct definition of the Mellin transform \eqref{Mellin} it may seem surprising that the large $s$ behaviour is controlled by a region ($t\sim -1$) which is far away from the integration contour. To argue that this is the case, we start by considering the inverse Mellin transform \eqref{inverseMellin} where the contour $\mathcal{C}$ is a straight line parametrized by $s=c+i\eta$ for some constant $2\D-\tilde \Delta_0<c<\tilde \Delta_0$ (the additional poles that are included in \eqref{psi0ancont} and \eqref{psiinfancont} for the analytic continuation will not affect this argument) and $\eta \in\mathbb{R}$. We take $t=e^{i\theta}$ and we integrate over~$\eta$  
\begin{equation}\label{inverseMellineta}
 f(e^{i \theta})=e^{i c \theta} \int_{-\infty}^{\infty} d\eta\, \Gamma(c+i \eta) \Gamma(2\Delta_{\phi}-c-i \eta)\, M(c+i\eta) \, e^{- \theta \eta} \, .
\end{equation}
We are interested in the behaviour of the integrand for $|\eta|\to \infty$. In this limit 
\begin{equation}\label{Gammaasyn}
 \Gamma(c+i \eta) \Gamma(2\Delta_{\phi}-c-i \eta) \sim e^{-\pi|\eta|}\, \eta^{2\Delta_{\phi}-1} \qquad |\eta|\to\infty \, .
\end{equation}
This means that the Gamma function prefactor accounts for the exponential behaviour \eqref{bound M} of $\hat{M}(s)$ for $|\text{Im}(s)|\to\infty$ predicted by the theorem in Section \ref{theorem}. This essentially motivates our choice of prefactor in \eqref{Mellin}. In particular, the exponential in \eqref{Gammaasyn} combined with that in \eqref{inverseMellineta} shows that the regime $\theta\to \pm\pi$ is controlled by the region $\eta\sim \mp\infty$. Let us make this more precise by defining
\begin{align}\label{Hdef}
 H(\eta)&\equiv\Gamma(c+i \eta) \Gamma(2\Delta_{\phi}-c-i \eta)\, M(c+ i \eta) \,e^{\pi|\eta|} 
\end{align}
so that the integral \eqref{inverseMellineta} can be rewritten as
\begin{equation}\label{Laplacetransf}
 f(e^{i \theta})=e^{i c \theta} \int_{0}^{\infty} d\eta \,H(-\eta) \, e^{-\eta(\pi-\theta)}+ e^{i c \theta} \int_{0}^{\infty} d\eta \,H(\eta) \, e^{-\eta(\pi+\theta)} \, .
\end{equation}
where we recognize two Laplace transforms of the functions $H(\pm\eta)$. A singular behaviour for $\theta=\pi$ originates from the first term in the sum \eqref{Laplacetransf}, while the singularity at $\theta=-\pi$ arises from the second term. More specifically, Tauberian theorems for the Laplace transform imply that for a function $H(\eta)\sim k\, \eta^{ \a}$ as $\eta\to \infty$ then 
\begin{equation}
\int_{0}^{\infty} d\eta\, H(\eta) \, e^{-\eta(\pi+\theta)}\sim k\, \Gamma(\a+1) (\theta+\pi)^{-\a-1}  \qquad \theta\to-\pi
\end{equation}
and similarly for the case $\theta\to\pi$. We are then led to the conclusion that the Regge behaviour \eqref{reggetheta} is reproduced by asking that
\begin{equation}
 H(\eta)\sim |\eta|^{2\Delta_{\phi}-1} \qquad |\eta|\to \infty \, .
\end{equation}
Combining this with \eqref{Hdef} and \eqref{Gammaasyn} we conclude that 
\begin{equation}
 M(c+i\eta)=\mathcal{O}(|\eta|^0) \qquad |\eta| \to \infty \, .
\end{equation}
Assuming that no Stokes phenomenon occurs for physical correlators we can extend this behaviour for any $\text{arg}(s)$ such that
\begin{equation}\label{resultasymM}
 M(s)=\mathcal{O}(|s|^0) \qquad |s|\to\infty \, .
\end{equation}
The absence of Stokes phenomenon is an assumption for which we do not have a proof. This assumption however is verified in all our examples and it was made also in the higher-dimensional case \cite{Penedones:2019tng}. 

We conclude this section with an important remark about the perturbative regime, which we will consider in Section \ref{sec:pert}. The result \eqref{resultasymM} is valid for the full nonperturbative Mellin amplitude. If the correlator contains a small parameter, it is often the case that order by order in the perturbative expansion the Regge behaviour is worse than in the full nonperturbative correlator\footnote{A typical example of this phenomenon is the function $\frac{1}{1-gz}$, which is regular for $z\to\infty$ but its expansion at small $g$ is more and more divergent.}. In Appendix~\ref{app:polesandseries} we  illustrate this in details in the context of the analytic sum rules discussed in the next section. In view of this aspect, it is then useful to formulate our result in a more general form. Let us consider a correlator $\tilde f(z)$ with a Regge behaviour
\begin{equation}
\tilde  f(z)=\mathcal{O}( z^{2\Delta_{\phi}+n}) \qquad z\to \frac12+i\infty
\end{equation}
for some positive integer $n$, then the associated Mellin amplitude will have a large $s$ asymptotics
\begin{equation}\label{Mlarges}
 M(s)=\mathcal{O}( |s|^n) \qquad |s|\to \infty \, .
\end{equation}


\section{Sum rules} 
\label{sumrules}

A common way to express the well-known fact that an arbitrary set of CFT data does not necessarily lead to a consistent CFT is through a set of sum rules for the CFT data. In the following, we will start from our definition of Mellin amplitude and derive an infinite set of sum rules. As we mentioned in the Introduction, these sum rules are not dispersive, according to the definition of \cite{Caron-Huot:2020adz}. This is essentially related to the behaviour at infinity obtained using our one-dimensional definition. In Section \ref{boundedness} we described how the product of Gamma functions in our definition \eqref{Mellin} leads to a nice behaviour for the Mellin amplitude $M(s)$ at $s=\infty$. However, the introduction of that prefactor leads also to the appearance of spurious poles in the integral \eqref{inverseMellin}. In a generic CFT, it is not expected that operators with the exact dimension $s=2\Delta_\phi+n$ are present in the spectrum. Thus, the poles of the Gamma functions must be compensated by zeros in the Mellin amplitude. This strategy was used in \cite{Penedones:2019tng,Carmi:2020ekr} to derive dispersive sum rules for the higher dimensional case, where the Mellin amplitude needs to have double zeros. Here, we will use the same idea to derive a new set of sum rules, which are characterized by single zeros of the Mellin amplitude. This makes these sum rules different and less powerful than the dispersive ones, but we believe that their derivation and the check of their validity on a set of known example provides an important consistency check of our results.

One may be concerned because the presence of single or double zeroes for the Mellin amplitude seems to be related to the choice of the prefactor in \eqref{Mellin}. This is actually not the case. The choice to factor out a prefactor in \eqref{Mellin} is related to having a nice polynomial behaviour for the function $M(s)$ at $s\to\infty$. If we were to pick a different prefactor (for instance using Gamma function squared, leading to double poles for the Mellin amplitude), the Mellin amplitude would contain an essential singularity at $s=\infty$ and this divergence would have to be compensated by the function $F_p(s)$, which we will use in \eqref{functional} to derive our sum rules. We can then safely conclude that the choice a prefactor is just a convenient trick, but it does not affect the resulting sum rules.

Finally, let us emphasize some important differences compared to the higher-dimensional strategy of the Mellin Polyakov bootstrap \cite{Polyakov:1974gs,Gopakumar:2016wkt,Gopakumar:2016cpb}. The derivation of the non-perturbative Polyakov consistency conditions used in \cite{Penedones:2019tng,Carmi:2020ekr} is quite subtle due to the presence of accumulation points in the twist spectrum of higher-dimensional CFTs. In our case, however, the situation is simpler. The twist accumulation points are related to the presence of a spin or, equivalently, to the need of introducing two Mandelstam variables. For us there is no spin and the only quantum number is the scaling dimension of the operators. Therefore, we do not expect any accumulation point in the spectrum and we will be able to impose the conditions \eqref{zeroes} without recurring to any analytic continuations.

\subsection{Properties of $M(s)$ and derivation of the sum rules}\label{derivationsumrules}

We start by summarizing the main properties of the Mellin amplitude $M(s)$ in~\eqref{Mellin}: 
\begin{itemize}
	\item $M$ is \textbf{crossing symmetric} 
	\begin{equation}
		M(s) = M(2\Delta_\phi-s) \, .
	\end{equation}
	\item $M$ has \textbf{poles} at the location of the physical exchanged operators in the two channels, i.e. $s=\D+k$ and $s=2\Delta_\phi-\D-k$ for $k\in \mathbb{N}  \, .$
	\item Generically, $M$ has single \textbf{zeros} compensating the poles of the prefactor 
	\begin{align}\label{zeroes}
		M(2\Delta_\phi+k) =0 \quad\text{and} \quad M(-k)=0 \quad \text{for} \quad k \in \mathbb{N} \,.
	\end{align}
	If the spectrum contains protected operators, some of these zeros might be absent.
	\item $M$ is \textbf{ bounded} for $|s|\to \infty$, see \eqref{resultasymM} .

	\item $M$ admits a crossing-symmetric \textbf{Mellin block expansion} 
	\begin{align}
		M(s) = \sum_\D \, c_\D\, M_\D(s) 
	\end{align}
	with $M_\D(s)$ given by the comparison with \eqref{hypergeom}.
\end{itemize}

The properties above will allow us to define a set of sum rules along the lines of \cite{Penedones:2019tng, Carmi:2020ekr}. 
Let $\omega_p$ be the functional 
\begin{align}\label{functional}
	\omega_{p_i} = \oint_{\mathbb{C}|_{\infty}} \frac{ds}{2\pi i}M(s)F_{p_i}(s)\, ,
\end{align}
where the contour here is a very large circle around infinity. When $F_{p_i}(s)$ is a sufficiently suppressed function at $s\to \infty$, we can take the limit of infinite radius for the circle and we get
\begin{align}
		\omega_{p_i} [M] =0	 \, .
\end{align}
For a nonperturbative Mellin amplitude characterized by the asymptotic behaviour \eqref{resultasymM} it is sufficient to ask that $F_{p_i}(s)\sim s^{-1-\e}$ for $\e>0$ as $|s|\to \infty$. As we mentioned at the end of Section \ref{boundedness}, when considering a perturbative expansion around GFF, the Regge behaviour may worsen and a sufficiently suppressed function $F$ would be required as detailed in Appendix \ref{app:polesandseries}.

The strategy to derive the sum rules simply consists in deforming the integration contour in \eqref{functional} to include all the poles of the integrand such that
 	\begin{align}\label{functionalwithM}
		\omega_{p_i} = \sum_{s^*} \text{Res}_{s=s^*}\left[M(s)\right]F_{p_i}(s^*)+\sum_{s^{**}} M(s^{**})\text{Res}_{s=s^{**}} \left[F_{p_i}(s)\right]=0 \, .
	\end{align}
	This equation already looks like a sum rule, but it depends on the value $M(s^{**})$ of the Mellin amplitude at the poles of $F_{p_i}(s)$. To avoid this issue one can simply choose $F_{p_i}(s)$ to have simple poles at the position of the zeros of $M(s)$. Therefore, we need a function $F_{p_i}(s)$ with poles at $s=-k$ or at $s=2\Delta_{\phi} +k$. Furthermore, the function $F_{p_i}(s)$ must not be crossing symmetric. Indeed, using the position of the poles in \eqref{leftpoles} and \eqref{rightpoles} and crossing symmetry for the residues \eqref{residues} we get
	\begin{align}\label{sumrulerightpoles}
		\omega_{p_i} = \sum_{s_R} \text{Res}_{s=s_R}(M(s))(F_{p_i}(s_R)-F_{p_i}(2\Delta_\phi-s_R)) \, ,
	\end{align}
so that any crossing symmetric function $F$ would lead to a trivial vanishing of $\omega_{p_i}$. Using the explicit expression for the residues \eqref{residues} we find the set of sum rules
\begin{align}\label{sumrulesgenericF}
	\sum_{\D,k} c_\D \frac{(-1)^{k+1}\Gamma(2\D)\Gamma(\D+k)}{\Gamma(\D)^2\Gamma(2\D+k)\Gamma(2\Delta_\phi-\D-k)\Gamma(k+1)}(F_{p_i}(\Delta+k)-F_{p_i}(2\Delta_\phi-\Delta-k))=0 \, .
\end{align}
A natural choice for the function $F$ is 
\begin{align}\label{Fp1p2}
 F_{p_1,p_2}(s)=\frac{1}{(s+p_1)(s+p_2)} \, ,\qquad p_1,p_2\in \mathbb{N} \, .
\end{align}
Notice that, despite the function $F_{p_1,p_2}(s)\sim \frac{1}{s^2}$ for $s\to \infty$, thanks to \eqref{sumrulerightpoles} only the crossing antisymmetric part of it matters, i.e. $F_{p_1,p_2}(s)-F_{p_1,p_2}(2\Delta-s)$ and one can easily check that this combination decays as $\frac{1}{s^3}$ for $s\to\infty$. Using this function we can derive the nonperturbative sum rules
\begin{align}\label{sumrulesnonpert}
	\sum_{\D,k} c_\D \tfrac{(-1)^{k+1}\Gamma(2\D)\Gamma(\D+k)}{\Gamma(\D)^2\Gamma(2\D+k)\Gamma(2\Delta_\phi-\D-k)\Gamma(k+1)}\tfrac{2(\Delta+k-\Delta_{\phi})(p_1+p_2+2\Delta_{\phi})}{(\Delta+k+p_1)(\Delta+k+p_2)(2 \Delta_{\phi}-\Delta-k+p_1)(2 \Delta_{\phi}-\Delta-k+p_2)}=0 \, .
\end{align}
Performing the sum over $k$ one obtains sum rules of the form 
\begin{align}
 \sum_{\D} c_\D  \a_{\Delta}=0
\end{align} \label{formsumrule}
with 
\begin{align}\label{alphasumrule}
\a_{\Delta}&=\frac{\Gamma(\Delta)}{\Gamma(2\Delta)\Gamma(2\Delta_{\phi}-\Delta)} \left(\mathcal{F}_{p_1,p_2}(\Delta)+\mathcal{F}_{-2\Delta_{\phi}-p_1,-2\Delta-p_2}(\Delta) \right) \, ,\\
 \mathcal{F}_{p_1,p_2}(\Delta)&=\tfrac{1}{p_1-p_2}\left(\tfrac{{}_3F_2(\Delta,p_1+\Delta,1+\Delta-2\Delta_{\phi};2\Delta,1+p_1+\Delta;1)}{(p_1+\Delta)}-\tfrac{{}_3F_2(\Delta,p_2+\Delta,1+\Delta-2\Delta_{\phi};2\Delta,1+p_2+\Delta;1)}{(p_2+\Delta)}\right)\, .
 \end{align}
As already mentioned in the Introduction, these sum rules differ from those found in \cite{Mazac:2018ycv,Ferrero:2019luz,Penedones:2019tng,Carmi:2020ekr}, which are dispersive sum rules having double zeros at the dimension (or twist in higher $d$) of double twist operators. Our functionals $\a_{\Delta}$ have single zeros at $\Delta=2\Delta_{\phi}+k$ for $k\in \mathbb{N}$ and $k\neq p_1,p_2$, implying that the functional changes sign at any of these zeros. The absence of any positivity property makes these sum rules less powerful and harder to use with the standard method of the modern conformal bootstrap. Despite this, we will now test them on some known examples and discuss their applicability in a perturbative setting. 

\subsection{Checks and applications}
Testing the sum rules \eqref{sumrulesnonpert} or  \eqref{sumrulesgenericF} on a fully nonperturbative spectrum is a task which is momentarily out of reach. Therefore, we start by testing it on the simplest possible $1d$ CFT, i.e. generalized free field theory, and perturbations thereof. 

\subsubsection{Generalized free theories} 
\label{subsubsec:GFFsumrules}
One may immediately raise several objections to our attempt of applying the sum rules \eqref{sumrulesgenericF} to GFF theories. First of all, as we discussed in Section \ref{sec:GFF} the definition of the Mellin amplitude for GFF requires the inclusion of a cut-off to regulate the exchange of the identity operator. Furthermore, a crucial assumption in the derivation of the sum rules was that $M(s)$ must have zeros for $s=-k$ and $s=2\Delta_{\phi}+k$, but GFF is precisely the example where two-particle operators with dimension $2\Delta_{\phi}+2n+k$ are exchanged and therefore those zero are absent. We will see that these issues can be avoided by considering the function
\begin{align} \label{GFFkernel}
 F_p(s)=\frac{1}{(2\Delta_{\phi}+2p-s)(2\Delta_{\phi}+2p-s+1)} \, .
\end{align}
Since this function has two poles and no residue at infinity we have the property
\begin{align}
 \text{Res}_{s=2\Delta_{\phi}+2p} (F_p(s))+ \text{Res}_{s=2\Delta_{\phi}+2p+1} (F_p(s))=0 \, .
\end{align}
Furthermore, one can easily check from \eqref{resultGFFepsilon} that $M_{\epsilon}^{\text{GFF}}(2\Delta_{\phi}+2p)=M_{\epsilon}^{\text{GFF}}(2\Delta_{\phi}+2p+1)$ for $p\in \mathbb{N}$. Combining these two properties, it is clear that the last term in \eqref{functionalwithM} vanishes even though $M(s)$ has no zeros at the positions of the poles of $F_p(s)$. We are then left with a sum over the residues of $M(s)$. For GFF, the position of the poles in principle depends on the regulator $\epsilon$, but the role of the regulator in \eqref{resultGFFepsilon} is to separate left and right poles. This is precisely what we have done to go from \eqref{functionalwithM} to \eqref{sumrulerightpoles}. Therefore, equation \eqref{sumrulesnonpert} can be used with $\e\to 0$ and, inserting the GFF spectrum $\Delta=2\Delta_{\phi}+2n$ we end up with the following sum rule
\begin{align}
	\sum_{n,k} c^{(0)}_n \tfrac{(-1)^{k+1}\Gamma(4(\Delta_{\phi}+n))\Gamma(2\Delta_{\phi}+2n+k)}{\Gamma(2\Delta_{\phi}+2n)^2\Gamma(4(\Delta_{\phi}+n)+k)\Gamma(-2n-k)\Gamma(k+1)}\tfrac{2 (2 \Delta_{\phi} +4 p+1) (\Delta_{\phi} +k+2 n)}{(k+2 n-2 p-1) (k+2 n-2
   p) (2 \Delta_{\phi} +k+2 n+2 p+1) (k+2 (\Delta_{\phi} +n+p))}=0 \,  .
\end{align}
Notice that the $\Gamma(-2n-k)$ factor in the denominator kills all the terms in this sum that are not compensated by a pole in the second ratio. The sum over $k$ then receives only two contributions at $k=2p-2n$ and $k=2p-2n-1$ which are present only for $p\geq n$. We are then left with a finite sum over $n$
\begin{align} \label{bosonicGFFsumrule}
	\sum_{n=0}^{p}c_n^{(0)}\frac{ \Gamma (2 p+1) \Gamma (4 (n+\Delta_{\phi} )) \Gamma (2 (p+\Delta_{\phi} )) \left(\Delta_{\phi} -2 n^2-4 \Delta_{\phi}  n+n+2 \Delta_{\phi}  p\right)}{\Gamma (2 (n+\Delta_{\phi} ))^2 \Gamma (2(p-n+1)) \Gamma (2 (n+p+2 \Delta_{\phi}+1 ))}=0 \, .
\end{align}
Clearly, each value of $p$ leads to an equation for $c_p^{(1)}$ in terms of all the other OPE coefficients with lower index. In other words, this sum rule can be solved recursively for $c^{(0)}_n$ determining everything in terms of $c^{(0)}_0$, which sets the overall normalization. Doing so, one easily finds that the unique solution to this sum rule is provided by \eqref{GFFOPE}. The same strategy can be used to determine the OPE coefficients for a fermionic GFF and for a GFF with $O(N)$ symmetry. We do this in Appendix \ref{otherGFF}.

\subsubsection{Perturbative sum rules}
In this section we further test our sum rules by using known CFT data for a class of perturbations around GFF. These perturbations, which will be treated in great details in Section \ref{sec:pert}, are constructed by introducing an effective field theory in AdS$_2$ background and considering the $1d$ boundary conformal field theory through the usual holographic dictionary. In particular, we will be interested in quartic contact interaction with derivatives. A classification of these independent contact interactions is given e.g. in~\cite{Mazac:2018ycv}, where the authors find there is a one-parameter family labelled by~$L$, where~$4L$ is the number of derivatives in the schematic interaction~$(\partial^L \Phi)^4$. As we mentioned at the end of Section \ref{boundedness}, single terms in the perturbative expansion of the correlator may have a worse Regge behaviour than the general bound \eqref{Reggebound}. In particular, let us consider a perturbed correlator 
\begin{equation}\label{pertcorr}
 f(t)=f^{\text{GFF}}(t)+g_{L} f_L^{(1)}(t)+O(g_L^2),
\end{equation}
where $L$ labels the maximum number of derivatives in the quartic interaction (i.e. the interaction term may involve a combination of terms with $\ell\leq L$ derivatives) and $g_L$ is the associated coupling. The Regge behaviour of this correlator is determined by the term with the maximum number of derivatives and it reads \cite{Mazac:2018ycv}
\begin{equation}
 f_L^{(1)}(z)\sim z^{2\Delta+2L-1} \qquad z \to \frac12+ i \infty.
\end{equation}
According to our discussion in Section \ref{derivationsumrules}, the associated Mellin amplitude will behave as
\begin{equation}
 M^{(1)}_L(s)\sim |s|^{2L-1} \qquad |s|\to \infty,
\end{equation}
and we need to choose a function $F_p(s)$ which vanishes at infinity faster than $|s|^{-2L}$. Here we will derive and check the sum rules for the cases $L=0$ and $L=1$. The strategy is the following. We use equation \eqref{sumrulesgenericF} to write down nonperturbative sum rules with a specific function $F_p(s)$ which will be chosen to decay sufficiently fast at $|s|\to \infty$ at a given value of $L$. We then expand the CFT data as
\begin{align}\label{expansionDelta}
 \Delta&=2 \Delta_{\phi}+ 2n+g_L \gamma_{L,n}^{(1)}+\mathcal{O}(g_L^2) \, ,\\
 c_{\Delta}&= c^{(0)}_n+ g_L c^{(1)}_{L,n} +\mathcal{O}(g_L^2)\label{expansionc}
\end{align}
and derive perturbative sum rules for $\gamma^{(1)}_{L,n}$ and $c^{(1)}_{L,n}$. We then check that these sum rules are satisfied by the $L=0,1$ results obtained in \cite{Mazac:2018ycv}, which read
\begin{align}\label{andimL0}
	\gamma_{0,n}^{(1)} &= \frac{\left(\frac{1}{2}\right)_n \left((\Delta_{\phi} )_n\right){}^2 \left(2 \Delta_{\phi} -\frac{1}{2}\right)_n}{(1)_n \,(2 \Delta_{\phi} )_n \left(\left(\Delta_{\phi} +\frac{1}{2}\right)_n\right){}^2}\, ,\\ \label{andimL1}
	\gamma_{1,n}^{(1)}&=A_{\Delta_{\phi}}^{-1} \gamma_{0,n}^{(1)}  \frac{2 n (4 \Delta_{\phi} +2 n-1)}{(\Delta_{\phi} +n-1) (2 \Delta_{\phi} +2 n+1)}( 16 \Delta_{\phi} ^5-13 \Delta_{\phi} ^3-3 \Delta_{\phi} ^2+16 \Delta_{\phi}  n^4+8 n^4+64 \Delta_{\phi} ^2 n^3\nonumber \\
&+16 \Delta_{\phi}  n^3-8 n^3+96 \Delta_{\phi} ^3 n^2+8 \Delta_{\phi} ^2 n^2-24 \Delta_{\phi}  n^2-2 n^2+64 \Delta_{\phi} ^4 n-28 \Delta_{\phi} ^2 n-2 \Delta_{\phi}  n+2 n)
\end{align}
where the result for $\gamma^{(1)}_{1,n}$ differs from \cite{Mazac:2018ycv} by an overall factor 
\begin{equation}
A_{\Delta_{\phi}} \equiv \frac{\Delta_{\phi}  (\Delta_{\phi} +1) (\Delta_{\phi} +2) (4 \Delta_{\phi}-1) (4 \Delta_{\phi} +1)^2 (4 \Delta_{\phi} +3)}{(2 \Delta_{\phi} +1)^2 (2 \Delta_{\phi} +3)}  \,  ,
\end{equation}
which we introduced to normalize the anomalous dimension as $\gamma_{1,1}^{(1)}=1$. Notice that $\gamma_{1,0}^{(1)}=0$. This is equivalent to a choice of basis for the set of independent interactions that can be built with up to one derivative. We will discuss this issue in details in Section \ref{sec:pert}. The OPE coefficients $c^{(1)}_n$ are given by the relation
\begin{equation}\label{OPEcoeff1}
 c^{(1)}_{L,n}=\partial_n(\gamma^{(1)}_{L,n} c^{(0)}_{n}) \, .
\end{equation}

\begin{figure}[htbp]
	\center
	\includegraphics[width=.9\linewidth]{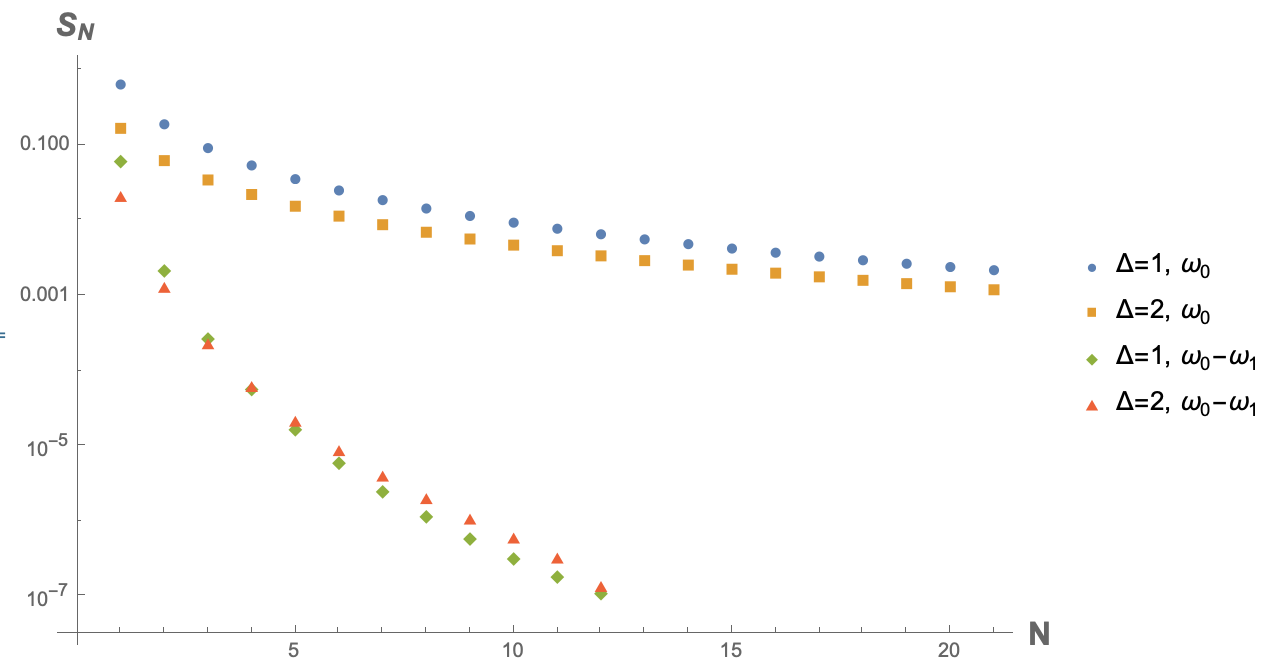}
		\caption{$L=0$ sum rules for external dimensions $\Delta_\phi=1,\Delta_\phi=2$ and different functionals $\omega_0$ and $\omega_0-\omega_1$. The plot shows the value of the finite sum truncated after $N$ terms in a logarithmic scale. The better convergence is seen for the $\omega_0-\omega_1$ functional, due to a better large $s$ behaviour.}\label{fig:sum-rule-phi4}
\end{figure}

\begin{figure}[htbp]
	\centering
	\includegraphics[width=0.9\linewidth]{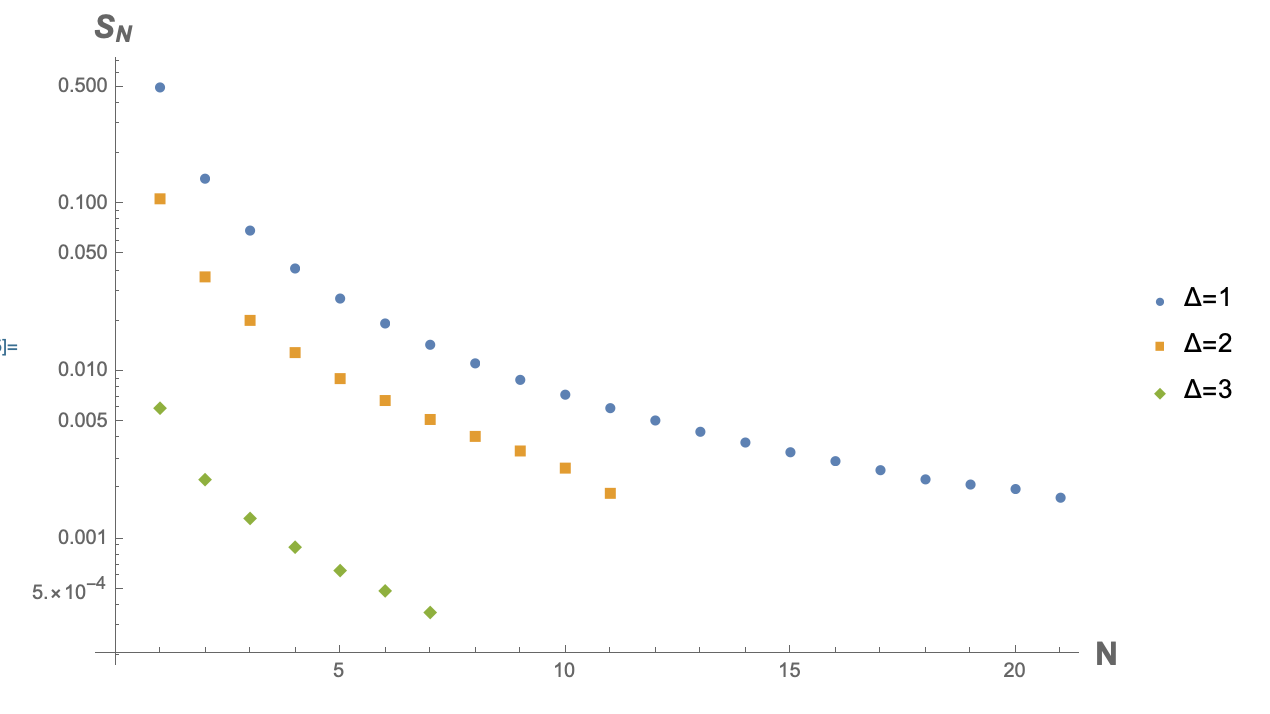}
	\caption{Sum rule applied for the $L=1$ case for $\Delta_\phi=1,2,3$, truncated after summing over N terms seen on a logarithmic scale. The slower convergence compared to $L=0$ is to be expected from the worse Regge behaviour.}
	\label{fig:sum-rule-dphi4}
\end{figure}

Let us start by the case $L=0$. In that case, the Regge behaviour is actually better than the GFF case, so we could even choose the function
\begin{equation}\label{functionforL0}
 F_p(s)=\frac{1}{s+p} \, ,
\end{equation}
which gives the sum rules
\begin{align}\label{sumruleL0}
	\sum_{\D,k} c_\D \frac{(-1)^{k+1}\Gamma(2\D)\Gamma(\D+k)}{\Gamma(\D)^2\Gamma(2\D+k)\Gamma(2\Delta_\phi-\D-k)\Gamma(k+1)(\D+k+p) (-2 \Delta +\D+k-p)}=0 \, .
\end{align}
Notice that we can also use the function \eqref{functionforL0} as a building block from which we can construct more suppressed functions of the class \eqref{Fp1p2}, for example
\begin{align}\label{combinationfunctions}
	F_{p,p+1}(s)=F_p(s)-F_{p+1}(s)=\frac{1}{(s+p)(s+p+1)} \, .
\end{align}
As we discussed below \eqref{Fp1p2}, the crossing antisymmetric part of this function decays as $\frac{1}{s^3}$ at large $s$ and for this reason we will also use it for the case $L=1$.
Inserting the expansions \eqref{expansionDelta} and \eqref{expansionc} into the sum rule \eqref{sumruleL0} we get two contributions: a finite sum from the terms where $k=p-2n$
\begin{align}\label{finitesum}
\tilde{\omega}_p&= \sum_{n=0}^{[\frac{p}{2}]}c_n^{(1)}\frac{\Gamma (p+1) \Gamma (4 (n+\Delta_\phi )) \Gamma (p+2 \Delta_\phi)}{(p-2 n)!\, \Gamma (2 (n+\Delta_\phi))^2 \Gamma (2 n+p+4 \Delta_\phi)}\\
&+c_n^{(0)}\gamma_n^{(1)} \frac{ \Gamma (p+1) \Gamma (4 (n+\Delta_\phi)) \Gamma (p+2 \Delta_\phi)}{2 (\Delta_\phi+p) (p-2 n)! \,\Gamma (2 (n+\Delta_\phi))^2 \Gamma (2 n+p+4 \Delta_\phi)} \eta(\Delta_\phi,n,p)  \, ,
\end{align}
where 
\begin{align}
	\eta(\Delta_\phi,n,p) =1+2 (\Delta_\phi +p) (-2 \psi_{2 n+2\Delta_\phi }+2 \psi_{4 n+4\Delta_\phi}-2 \psi_{ 2 n+p+4 \Delta_\phi }+\psi _{p+2 \Delta_\phi}+\psi_{p+1}) 
\end{align}
with $\psi_n=\frac{\Gamma'(n)}{\Gamma(n)}$, and an infinite sum 
\begin{equation}
 \tilde{\omega}'_p=\sum_{n,k\neq p-2n}c_n^{(0)} \gamma_n^{(1)} \frac{2 (k+2 n)! (\Delta_\phi +k+2 n) \Gamma (4 (n+\Delta_\phi )) \Gamma (k+2 n+2 \Delta_\phi )}{k! (k+2 n-p) \Gamma (2 (n+\Delta_\phi ))^2 \Gamma (k+4 (n+\Delta_\phi )) (2 \Delta_\phi +k+2 n+p)} \, .
\end{equation}
This last term does not allow us to follow the same strategy we used for GFF to extract the value of $\gamma^{(1)}_{L,n}$. This is consistent with the fact that these sum rules are not as constraining as in the GFF case and we may expect more than one solution. Therefore, instead of using the sum rules to derive the CFT data, we limit ourselves to numerically check the validity of the equation
\begin{equation}
 \omega_p=\tilde{\omega}_p+\tilde{\omega}'_p=0
\end{equation}
by inserting the data \eqref{andimL0}, \eqref{andimL1} and \eqref{OPEcoeff1}. In figure \ref{fig:sum-rule-phi4} we show our results for the case $p=0$ both for the functional $\omega_0$ and for the combination \eqref{combinationfunctions}, i.e. the functional $\omega_0-\omega_1$. It is very clear from the plot that the latter shows a faster convergence reflecting the better asymptotic behaviour at $|s|\to \infty$.  In Figure \ref{fig:sum-rule-dphi4} we show the analogous plot for the $L=1$ case, where we used the function \eqref{combinationfunctions} since we needed a large $s$ behaviour at least $\frac{1}{s^3}$.

We conclude with a couple of remarks. Since the finite sum in \eqref{finitesum} always ranges up to $[\frac{p}{2}]$, we can always find a combination of functionals $\omega_k$ such that the contributions proportional to $c^{(1)}_{L,n}$ are all cancelled and we can write down a sum rule which only involves the anomalous dimensions $\gamma^{(1)}_{L,n}$
\begin{equation}\label{New sum rule}
\sum_{k=0}^{2p+1}\frac{(-1)^{1+k}\,\Gamma (2 p+2)}{\Gamma (k+1) \,\Gamma (2 p+3-k)} \,\omega_k=0 \, .
\end{equation}
The simplest case is $p=0$ where \eqref{New sum rule} simply gives $\omega_0-2\omega_1=0$. Even after doing that, however, we are left with an infinite sum with alternating sign, which we did not manage to use constructively to extract a solution. Nevertheless, notice that, once the anomalous dimensions are known we can use our sum rules at fixed $p$ to determine recursively the value of the OPE coefficients $c^{(1)}_{L,n}$ following the same strategy as in the GFF case. This provides a very convoluted way to rederive the relation \eqref{OPEcoeff1}. 

\section{Perturbative results}
\label{sec:pert}

In this section we consider deformations from generalized free-field theory produced by effective interactions in a bulk AdS$_2$ field theory. In this holographic AdS$_2$/CFT$_1$ setup the background AdS$_2$ metric is not dynamical, corresponding to the absence of a stress tensor in the boundary CFT$_1$. According to the usual dictionary, a massive free scalar field $\Phi$ in AdS$_2$ is dual to a boundary $1d$ generalised free field $\phi$. We deform this theory by a quartic self-interactions with an arbitrary number $L$ of derivatives 
\be\label{phi4lagrangian}
S=\int dx dz\,\sqrt{g}\,\big[ \,g^{\mu\nu}\,\partial_\mu\Phi\,\partial_\nu\Phi+m^2_{\Delta_\phi}\Phi^2+g_L\,(\partial^L\Phi)^4\,\big]\,,\qquad L=0,1,\dots
\ee
where we use the AdS$_2$ metric in Poincar\'e coordinates $ds^ 2=\frac{1}{z^2}(dx^ 2+dz^ 2)$. The mass $m^2_{\Delta_\phi}=\Delta_\phi(\Delta_\phi-1)$ is fixed in units of the AdS radius so that $\Delta_\phi$ is the dimension, independent of $g_L$, of the field $\Phi$ evaluated at the boundary, $\phi(x)$\footnote{When we introduce an interaction, such as \eqref{phi4lagrangian}, there will be Witten diagrams contributing to the mass renormalization of $\Phi$. We can always choose the bare mass in such a way that the dictionary is preserved and $\Delta_{\phi}$ is not modified.}.  We will limit our analysis to tree-level correlators, and thus consider only contact diagrams, whose building blocks are the $D$-functions~\cite{Liu:1998ty,DHoker:1999kzh, Dolan:2003hv}  reviewed in Appendix~\ref{app:Dfunctions}. 
The writing $(\partial^L\Phi)^4$ above is symbolic, denoting a complete and independent set of quartic vertices with four fields and up to $4L$ derivatives\footnote{The fact that a complete and independent basis of vertices is labelled by 1/4 the number of derivatives can be seen using integration by parts and the equations of motion, or noticing that the counting of physically  distinct four-point interactions is equivalent to the counting of crossing-symmetric polynomial $S$-matrices in 2D Minkowski space, see discussion in~\cite{Mazac:2018ycv, Ferrero:2019luz}.}. In the following, we will present a particularly convenient basis for these interactions, which will allow us to derive a closed-form expression for the tree-level correlator in Mellin space. Consider the interaction Lagrangian
\begin{equation}\label{interactionlag}
 \mathcal{L}_{L}=g_{L}\left[\prod_{k=0}^{L-1}\left(\tfrac{1}{2}\partial_\mu \partial^\mu-(\Delta_{\phi}+k)(2(\Delta_{\phi}+k)-1)\right)\Phi^2 \right]^2 \, .
\end{equation}
This looks like a very complicated term, but it contains four fields $\Phi$ and $4L$ derivatives, so by the argument above it must be effectively a linear combination of operators like $(\partial^{\ell}\Phi)^4$ for $\ell\leq L$. The advantage of this interaction is that the corresponding correlator computed via Witten diagrams reads 
\be\label{ansatzcorrelator}
f^{(1)}_L(z) = \frac{4^{L-1}\pi^{-\frac32}\Gamma(2\Delta_{\phi}-\frac{1}{2}+2L)}{\Gamma(\Delta_{\phi}+\frac12)^4}  z^{2\Delta_\phi}\,(1+z^{2L}+(1-z)^{2L})\bar{D}_{\Delta_\phi+L, \Delta_\phi+L, \Delta_\phi+L, \Delta_\phi+L}(z) \, ,
\ee
where the $\bar{D}$-functions are introduced in Appendix \ref{app:Dfunctions}. If one starts with some specific $4L$-derivative interaction, such as $(\partial^{L}\Phi)^4$, the explicit computation through Witten diagrams shows the appearance of several other combinations of $D$-functions with different weights (we will perform some of these computations explicitly in Section \ref{diagrammaticchecks}). Nevertheless, by the argument above these results cannot be independent of those obtained using $\mathcal{L}_{L}$ and therefore the result must be expressible as a linear combination $\sum_\ell a_{\ell}f^{(1)}_\ell(z)$. This requires a series of non-trivial identities among $\bar D$ functions, some of which we derive in Section \ref{diagrammaticchecks}. 

Using \eqref{ansatzcorrelator} as a basis for $4L$-derivative results, we can take its Mellin transform. The first step is to compute the Mellin transform of the function $\bar{D}_{\Delta_\phi\Delta_\phi\Delta_\phi\Delta_\phi}(t)$. In this section we consider the reduced Mellin amplitude $\hat{M}(s)\equiv M(s) \Gamma(s)\Gamma(2\Delta_{\phi}-s)$ and we need to compute
\be\label{MellinDbar}
\hat{M}_{\Delta_\phi}(s) = \int_0^\infty dt \,\bar D_{\Delta_\phi\Delta_\phi\Delta_\phi\Delta_\phi}(t)  \, \Big(\frac{t}{1+t}\Big)^{2\Delta_\phi}\,t^{-1-s}   \, .
\ee
A closed-form expression for the $\bar D$ functions is not available and dealing with integral representations is quite hard. Therefore, we considered the case of integer $\Delta_{\phi}$, where simple explicit expressions for the $\bar D$ functions are known (see~\eqref{Dbar-explicit}-\eqref{Dbar-explicit-end}) and we inferred the general form
\begin{align}\label{MellinL0}
	\hat{M}_{\Delta_\phi}(s) &=\pi \csc(\pi s)\,\Big(  \pi \cot(\pi s)P_{\Delta_\phi}(s) -\sum_{k=1}^{2\Delta_\phi-1}\frac{P_{\Delta_\phi}(k)}{s-k} \Big)\, ,\\
		P_{\Delta_\phi}(s) &= 	2 \sum_{n=0}^{\Delta_{\phi}-1}   (-1)^{n} \frac{\Gamma(2 n + 1) \Gamma^4(\Delta_{\phi}) \Gamma(\Delta_{\phi}+n)}{\Gamma^4(n + 1)\Gamma(\Delta_{\phi}-n)\Gamma(2(\Delta_{\phi}+n))} (2\Delta_{\phi}-s)_n(s)_n\,,
\end{align}
The functions $P_{\Delta_\phi}(s)$ are effectively just polynomials of order $2\Delta_\phi-2$. Defining \be
Q_{\Delta_{\phi}} (s(s-2\Delta_{\phi}))\equiv P_{\Delta_{\phi}}(s) \, ,
\ee
we have, for the first few cases
\begin{align} \label{DfunctMellintable}
{\renewcommand\arraystretch{1.3} 
\begin{tabular}{ c | c }
$\Delta_{\phi}$ &  Q$_{\Delta_{\phi}} (x)$ \\
\hline 
1  &  2 		\\																										
2  & $\frac{1}{15} (5 + x) $ 	\\																					
3  & $\frac{1}{315} (84 + 17 \,x + x^2) $		\\											
4  & $\frac{1}{30030} (15444 + 2889 \,x + 206\, x^2 + 5\, x^3)$	\\
5  & $\frac{1}{765765} (1400256 + 239640\, x + 17387\, x^2 + 570\, x^3 + 7\, x^4)$	\\
\end{tabular}      }   
\end{align}

The functions $P_{\Delta_\phi}(s)$ can also be rewritten as
\begin{align}\label{PDelta}
		P_{\Delta_\phi}(s) &= 	2 \frac{\Gamma(\Delta_\phi)^4}{\Gamma(2\Delta_\phi)}	{}_4{F}_3(\{\textstyle\frac{1}{2},s,1-\Delta_\phi,2\Delta_\phi-s\};\{1,1,\Delta_\phi+\frac{1}{2}\};1)\, ,
\end{align}
 Notice the important fact that in cross-ratio space a closed-form expression for the $\bar D$ functions is not known, while in Mellin space it looks reasonably simple, at least for integer $\Delta_{\phi}$. This is similar to what happens in the higher dimensional case, where this occurrence is even more striking as the reduced Mellin transform of the $\bar D$ functions is simply a product of Gamma functions\footnote{It is often said that the Mellin transform of contact interactions is one, but this assumes that the correct product of Gamma function has been factored out~\cite{Mack:2009mi,Penedones:2010ue}.}. In the one-dimensional case, we could not find such a simple representation for the contact interactions, but the fact we could write the result in a closed form is already a notable improvement compared to cross-ratio space and as we will see, it will allow us to successfully extract new CFT data. Furthermore, in Section \ref{sec:alternativeMellin} we will present an alternative definition of Mellin transform which leads to simpler results for the contact interaction.  

Knowing the Mellin transform for the $\bar D$ functions, it is simple to compute the Mellin transform of \eqref{ansatzcorrelator}
\begin{align}\label{ML}
	\hat{M}^{(1)}_{L}(s)&=\int_0^\infty dt \, f^{(1)}_L(t)\, \Big(\frac{t}{1+t}\Big)^{2\Delta_\phi}\,t^{-1-s}=  \,\sum_{k=0}^{2L}	c_{k,L} \, \hat M_{\Delta_\phi+L}(s+k)\, ,\\
2 c_{k,L} &= \frac{\Gamma(2L+1)}{\Gamma(k+1)\Gamma(2L-k+1)}+\delta_{k,0}+\delta_{k,2L}\, .
\end{align}
Notice that the presence of double poles for integer values of $s$ in~\eqref{MellinL0} is not in contradiction with the general single-pole structure of the nonperturbative Mellin amplitude~\eqref{Mellinpoles}-\eqref{psiinfsum}, but just a consequence of the perturbative expansion of those single poles 
at this first order of perturbation theory as detailed in Appendix \ref{app:polesandseries}. Moreover, the structure of~\eqref{MellinL0}  is such that both single and double poles cancel (as they should) within the region of convergence~\eqref{convergence} of the integral~\eqref{MellinDbar}, which in this case ($\Delta_0=2\Delta_\phi$) is $0<\text{Re}(s)<2\Delta_\phi$.  The cancellation of the double poles is evident, given the poles of $\cot(\pi s)$ and the explicit poles in the sum.  The cancelling of the single poles stems  from a property of the polynomial $P_{\Delta_\phi}(s)$, which ensures the cancellation of the finite part in the expression multiplying $\csc(\pi s)$ when expanded  around integer values of~$s$, $0<s<2\Delta_\phi$. 
As we will see, this structure is consistent with the OPE expansion.

We stress that equation \eqref{ML} is a closed-form expression for the first-order perturbation around GFF generated by a quartic interaction with any number of derivatives. Under the assumption that the deformation from GFF described by these interactions only modifies two-particle data, see~\eqref{pertdelta}-\eqref{pertOPE} below, one can extract these data. In particular, in Section \ref{CFTdata} we will show that the anomalous dimension of two-particle operators receives the following correction
\begin{equation}\label{expansiondim}
 \Delta=2 \Delta_{\phi}+2 n+g_{L} \hat\gamma_{L,n}(\Delta_{\phi})
\end{equation}
with
\begin{align}
 \hat{\gamma}_{ L, n}^{(1)}(\Delta_{\phi})&=\frac{\Gamma(L+\Delta_{\phi})^4}{\Gamma(2L+2\Delta_{\phi})}\sum_{p=0}^{2n} \sum_{k=2L-p}^{2L}\sum_{l=0}^{k+p-2L}(-1)^k c_{k,L} \times\label{greatresult}\\   &\frac{(4\Delta_{\phi}+2n-1)_p (-2n)_p (2L-k-p)_l (1-\Delta_{\phi}-L)_l (2\Delta_{\phi}+k+p)_l (\tfrac12)_l}{(l!)^3 (2\Delta_{\phi})_p (\tfrac12+\Delta+L)_l} \nonumber  \, .
\end{align}
In order to compare these results with those computed with bootstrap methods in~\cite{Mazac:2018ycv, Ferrero:2019luz} for $L=0,1,2,3$, we have to change basis in the space of couplings. Since the bootstrap approach is blind to the specific values of the couplings $g_L$ in \eqref{expansiondim}, one needs to establish a criterium to organize the set of independent data. The criterium that is used in~\cite{Mazac:2018ycv, Ferrero:2019luz} consists in setting
\begin{equation}\label{condvanandim}
 \gamma_{L,n}(\Delta_{\phi})= 0 \qquad n<L \, .
\end{equation}
In our approach this is implemented by taking a linear combination
\begin{equation}\label{lincombandim}
\gamma_{L,n} =\sum_{\ell=0}^{L} a_{\ell} \hat\gamma_{\ell,n}
\end{equation}
and fixing the $L+1$ $a_\ell$ coefficients in~\eqref{lincombandim}, using the $L$ conditions \eqref{condvanandim} and the normalization $\gamma_{L,L}(\Delta_{\phi})=1$. Following this strategy in Section \ref{CFTdata}, we will reproduce the known results for $L\leq 3$ and present new results for $L\leq 8$ at any $\Delta$ and $n$. We stress however that equation \eqref{greatresult} is valid for any $L$, so, up to the algorithmic procedure of fixing the $a_{\ell}$ coefficients, one can easily extract the result for any given $L$.

%
Below we will show how the interaction \eqref{interactionlag} leads to the correlator~\eqref{ansatzcorrelator} through explicit Witten diagrammatics. We will also see, for the cases $L=0,1,2$, how other interaction terms lead to results that can be rearranged as linear combinations of the eigenfunctions \eqref{ansatzcorrelator}. We will then proceed to  extract CFT data in the bootstrap normalization.

 
\subsection{Diagrammatics} \label{diagrammaticchecks}

Here we consider the interaction Lagrangian \eqref{interactionlag} and we show that it leads to the correlator \eqref{ansatzcorrelator} using Witten diagrams. The result of the Wick contractions is
\begin{multline}\label{wickforspecialint}
 \braket{\phi(x_1)\phi(x_2)\phi(x_3)\phi(x_4)}^{(1)} =\\ \sum_{\text{perms}} g_L \int \frac{dx dz}{z^2} \mathcal{D} \left(K_{\Delta_\phi}(x,z;x_1) K_{\Delta_{\phi}}(x,z;x_2)  \right) \mathcal{D} \left(K_{\Delta_\phi}(x,z;x_3) K_{\Delta_{\phi}}(x,z;x_4)  \right)\, ,
\end{multline}
where we defined
\begin{align}\label{Eq: D basis}
\mathcal{D}= \prod_{k=0}^{L-1}\left(\frac{1}{2}\partial_{\mu}\partial^{\mu}-(\Delta_{\phi}+k)(2(\Delta_{\phi}+k)-1)\right) 
\end{align}
acting on the bulk point and we used the bulk-to-boundary propagator with the conventions of \eqref{bulktoboundary}.
Using recursively the identity
\begin{align}
	-2x_{ij}^2\Delta^2 \tilde K_{\Delta+1}(x,z;x_i) \tilde K_{\Delta+1}(x,z;x_j) = (\tfrac{1}{2}\partial_{\mu}\partial^{\mu}-\Delta(2\Delta-1))(\tilde K_{\Delta}(x,z;x_i) \tilde K_{\Delta}(x,z;x_j)) \, ,
\end{align}
which can be derived from \eqref{identityder}, we obtain
\begin{align}
\mathcal{D} \left(\tilde K_{\Delta_\phi}(x,z;x_1) \tilde K_{\Delta_{\phi}}(x,z;x_2)  \right) = (-2x_{12}^2)^L\left(\frac{\Gamma(\Delta_{\phi}+L)}{\Gamma(\Delta_{\phi})}\right)^2 \tilde K_{\Delta_\phi+L}(x,z;x_1) \tilde K_{\Delta_\phi+L}(x,z;x_2) \, .
\end{align}
Inserting this into equation \eqref{wickforspecialint}, summing over the permutations and remembering the definition of $\mathcal{C}_{\Delta}$ in \eqref{bulktoboundary}, we get 
\begin{multline}
	\braket{\phi(x_1)\phi(x_2)\phi(x_3)\phi(x_4)}^{(1)}_L = g_L [(x^2_{13}x^2_{24})^{L}+(x^2_{12}x^2_{34})^{L}+(x^2_{14}x^2_{23})^{L}] \times \\
	2^{2L-1} \pi^{-2}\left(\frac{\Gamma(\Delta_{\phi}+L)}{\Gamma(\Delta_{\phi}+\frac12)}\right)^4  
	D_{\Delta_{\phi}+L,\Delta_{\phi}+L,\Delta_{\phi}+L,\Delta_{\phi}+L}(x_1,x_2,x_3,x_4) \, .
\end{multline}
Using \eqref{Dbar} we immediately get
\begin{multline}\label{correlatorL}
\braket{\phi(x_1)\phi(x_2)\phi(x_3)\phi(x_4)}^{(1)}_L=g_L (1+z^{2L}+(1-z)^{2L})\times\\ \frac{4^{L-1}\pi^{-\frac32}\,\Gamma(2\Delta_{\phi}-\frac{1}{2}+2L)}{\Gamma(\Delta_{\phi}+\frac12)^4\,(x^2_{13}\,x^2_{24})^{\Delta_{\phi}}}  \bar{D}_{\Delta_\phi+L, \Delta_\phi+L, \Delta_\phi+L, \Delta_\phi+L}(z)
\end{multline}
in perfect agreement with \eqref{ansatzcorrelator}. In the following, we will perform some diagrammatic checks on the claim that, starting for an arbitrary interaction with up to $L$ derivatives, the result can be recast as a linear combination of $f^{(1)}_{\ell}(z)$ with $\ell\leq L$.
 
\subsubsection{Checks}

The case with no derivatives is trivial since the $D$-function are defined precisely as the first-order correlator with a quartic contact interaction with no derivatives
\begin{align}
\braket{\phi_{\D_\phi}(x_1)\phi_{\D_\phi}(x_2)\phi_{\D_\phi}(x_3)\phi_{\D_\phi}(x_4)}^{(1)}_{L=0}&= g_0\,4!\,\mathcal{C}^ 4_{\D_\phi}\, D_{\Delta_\phi\Delta_\phi\Delta_\phi\Delta_\phi}(x_1,x_2,x_3,x_4)\\\label{corrL0}
&=g_0\,\frac{3 \pi^{-\frac32}\,\Gamma(2\D_\phi-\frac{1}{2})}{4\,\Gamma(\D_\phi+\frac12)^ 4 (x^2_{13}\,x^2_{24})^{\D_\phi}}\,  {\bar D}_{\Delta_\phi\Delta_\phi\Delta_\phi\Delta_\phi}(z)\,, \nonumber
\end{align}
in agreement with \eqref{correlatorL} for $L=0$. 

Let us consider now the interaction term 
\begin{align}
	\mathcal{L}_{\textrm{int}} = -(g^{\mu \nu}\partial_\mu \phi\partial_\nu \phi) \, .
\end{align}
Using  the identity~\eqref{identityder}, the  result of the Wick contractions can be written explicitly in terms of $\bar{D}$-functions as
\begin{align}\nonumber
&	\braket{\phi_{\D_\phi}(x_1)\phi_{\D_\phi}(x_2)\phi_{\D_\phi}(x_3)\phi_{\D_\phi}(x_4)}_{L=1}^{(1)}=  \frac{\mathcal{C}_{\Delta_\phi}^4}{(x_{13} x_{24})^{2\Delta_\phi}}\frac{\sqrt{\pi}\Gamma(2\Delta_\phi-\frac{1}{2})}{\Gamma(\Delta_\phi)^4} (3\Delta_\phi^4\bar{D}_{\Delta_\phi \Delta_\phi \Delta_\phi \Delta_\phi}\\\nonumber
	&\qquad -4\Delta_\phi^2\left(2\Delta_\phi-\frac{1}{2}\right)(\bar{D}_{\Delta_\phi \Delta_\phi \Delta_\phi+1 \Delta_\phi +1}+\bar{D}_{\Delta_\phi \Delta_\phi+1 \Delta_\phi \Delta_\phi +1}+\bar{D}_{\Delta_\phi+1 \Delta_\phi \Delta_\phi \Delta_\phi +1})\\
	&\qquad +4\left(2\Delta_\phi-\frac{1}{2}\right)\left(2\Delta_\phi+\frac{1}{2}\right)\left(1+z^2+(1-z)^2\right)\bar{D}_{\Delta_\phi+1 \Delta_\phi+1 \Delta_\phi+1 \Delta_\phi +1}) \, .
\end{align}
Using~\eqref{Dbarconseq} we can then rewrite this as 
\begin{align}\label{contactL1}
\braket{\phi_{\D_\phi}(x_1)\phi_{\D_\phi}(x_2)\phi_{\D_\phi}(x_3)\phi_{\D_\phi}(x_4)}_{L=1}^{(1)}=  \frac{1}{(x_{12}\, x_{34})^{2\Delta_\phi}}\,\big( \,a_0\,f_{0}^{(1)}(z)+a_1\,f^{(1)}_1(z)\big)
\end{align}
with
\begin{align}
a_0&=-\frac14 \Delta_{\phi}^3\, (5\Delta_{\phi}-2)\, , & a_1&= \frac{1}{4}\, .
\end{align} 
The corresponding Mellin amplitude simply reads
\begin{align}\label{MellinL1}
	\hat M_{L=1}^{(1)}(s)&= a_0 \, \hat M_{\Delta_\phi}(s)+ a_1\,\big[\,\hat M_{\Delta_\phi+1}(s)+ \hat M_{\Delta_\phi+1}(s+1)+\hat M_{\Delta_\phi+1}(s+2)\,\big]\,,
\end{align}
where $\hat M_{\Delta_{\phi}}(s)$ is given in \eqref{MellinL0}.

 In the case $L=2$,  we consider the explicit interaction term
\begin{align}
	\mathcal{L}_{\textrm{int}} = -g\,(g^{\mu \nu}\partial_\mu \phi\partial_\nu \phi)^2\, .
\end{align}
Once again, the result of the Wick contractions can be expressed in terms of $\bar D$ functions (some details on this procedure are outlined in Appendix \ref{WittenL2}) and we can rewrite it as
\begin{align}\label{lincombL2}
&\braket{\phi_{\D_\phi}(x_1)\phi_{\D_\phi}(x_2)\phi_{\D_\phi}(x_3)\phi_{\D_\phi}(x_4)}_{L=2}^{(1)}= \frac{1}{(x_{13}\,x_{24})^{2\Delta_\phi}} \left(a_0 \,f_0^{(1)}(z)+a_1 \,f_1^{(1)}(z)+a_2\,f_2^{(1)}(z)\right)
\end{align}
with
\begin{align}
	a_0&=-\frac{7}{2}\Delta_{\phi} +2\Delta_{\phi}^2 +55\Delta_{\phi}^3-22\Delta_{\phi}^4-21\Delta_{\phi}^5\, ,\\
	a_1&=\frac{3}{2}+7\Delta_{\phi}+14\Delta_{\phi}^2+14\Delta_{\phi}^3+8\Delta_{\phi}^4\, ,\\
	a_2&=1\, .
\end{align}
Notice that, unlike the $L=1$ case, we found that the identities listed in Appendix \ref{app:Dfunctions} are not enough to recombine the result in the form \eqref{lincombL2}, but we need to use a new relation
\begin{align}\label{Eq: New relation}
	z^2\bar{D}_{\Delta_\phi+1\Delta_\phi+1\Delta_\phi+2\Delta_\phi+2}+\bar{D}_{\Delta_\phi+1\Delta_\phi+2\Delta_\phi+1\Delta_\phi+2}+(1-z)^2\bar{D}_{\Delta_\phi+2\Delta_\phi+1\Delta_\phi+1\Delta_\phi+2}=\nonumber \\\frac{\Delta_\phi^4}{2(3+4\Delta_\phi)}\bar{D}_{\Delta_\phi\Delta_\phi\Delta_\phi\Delta_\phi}+\frac{3+8\Delta_\phi+6\Delta_\phi^2}{2(3+4\Delta_\phi)}\left(1+z^2+(1-z)^2\right)\bar{D}_{\Delta_\phi+1\Delta_\phi+1\Delta_\phi+1\Delta_\phi+1} \, ,
\end{align}
which we tested numerically and for integer values of $\Delta_\phi$. This identity is not a consequence of the higher-dimensional ones given in \cite{Dolan:2000ut} and we believe it is a new, inherently one-dimensional relation among $D$ functions. Studying other interactions would lead to new identities of this kind.

The Mellin amplitude associated to \eqref{lincombL2} can be again put in the form~\eqref{ML}, and reads explicitly
	\begin{align}
	\!\!\! & \hat{M}_{L=2}^{(1)}(s)= a_0 \hat{M}_{\Delta_\phi}(s)+2 a_1 (\hat{M}_{\Delta_\phi+1}(s)+\hat{M}_{\Delta_\phi+1}(s+1)+\hat{M}_{\Delta_\phi+1}(s+2))\nonumber \\
	&+2 a_3 \left(\hat{M}_{\Delta_\phi+2}(s)+2\hat{M}_{\Delta_\phi+2}(s+1)+3\hat{M}_{\Delta_\phi+2}(s+2)+2\hat{M}_{\Delta_\phi+2}(s+3)+\hat{M}_{\Delta_\phi+2}(s+4)\right) \, .
\end{align}

\subsection{CFT data}\label{CFTdata}

Given the closed-form expression \eqref{ML} for the perturbative Mellin amplitude, we can use it to extract the CFT data
\begin{align}\label{pertdelta}
\Delta&\equiv \D_{n,L}=2\D_\phi+2n+g_L\, \hat{\gamma}_{ L, n}^{(1)}(\Delta_{\phi})+\dots\, ,\\\label{pertOPE}
c_{\D}&\equiv c_{n,L}=c_n^{(0)}(\Delta_{\phi})+g_L\,c_{L,n}^{(1)}(\Delta_\phi)+\dots \,,
\end{align}
where $n\in \mathbb{N}$ and $c_n^{(0)}$ is given in~\eqref{GFFOPE}. In~\eqref{pertdelta}-\eqref{pertOPE} we are assuming that the AdS interaction only modifies the CFT data of the two-particle operators exchanged in GFF. If we insert~\eqref{pertdelta}-\eqref{pertOPE} in the general Mellin OPE expansion~\eqref{Mellinpoles}, at first order in $g_L$  one obtains double and single poles at $s=2\Delta_{\phi}+p$,  $p\in \mathbb{N}$. One can then compare the corresponding residues with the ones in the tree-level Mellin amplitude~\eqref{ML}, which amounts to solve the equations formally written as
\begin{align}\label{doublepoles}
 &\sum_{n}  c_n^{(0)}(\Delta_{\phi}) \hat{\gamma}_{ L, n}^{(1)}(\Delta_{\phi})\,{\textstyle{\frac{(-1)^p \Gamma(4\Delta_{\phi}+4n)\Gamma(2\Delta_{\phi}+p)^2}{\Gamma(2\Delta_{\phi}+2n)^2\Gamma(4\Delta_{\phi}+2n+p)\Gamma(p-2n+1)}}}\,
=\lim_{s\rightarrow 2\Delta_{\phi}+p}(s-2\Delta_{\phi}-p)^2 \hat{M}^{(1)}_L(s) \,,
\\\label{singlepoles}
&\sum_{n}(c_{L,n}^{(1)}(\Delta_{\phi})\!+c_{n}^{(0)}(\Delta_{\phi}) \hat{\gamma}_{ L, n}^{(1)}(\Delta_{\phi})\partial_n)\,{\textstyle{\frac{(-1)^p \Gamma(4\Delta_{\phi}+4n)\Gamma(2\Delta_{\phi}+p)^2}{\Gamma(2\Delta_{\phi}+2n)^2\Gamma(4\Delta_{\phi}+2n+p)\Gamma(p-2n+1)}}}\,
=\text{Res}_{s= 2\Delta_{\phi}+p} \hat{M}^{(1)}_L(s)
\end{align}
for the leading-order corrections $\hat{\gamma}_{ L, n}^{(1)}(\Delta_{\phi})$ and $c_{L,n}^{(1)}(\Delta_{\phi})$.  Since the function $\hat{M}^{(1)}_L(s)$ is known explicitly, it is possible to write down a linear system for the anomalous dimensions $\hat{\gamma}_{ L, n}^{(1)}(\Delta_{\phi})$. To do this, let us use the form \eqref{MellinL0} for $\hat{M}^{(1)}_{\Delta_{\phi}}(s)$ to rewrite \eqref{doublepoles}  as
\begin{align}\label{system}
 &\sum_{n}  c_n^{(0)}(\Delta_{\phi}) \hat{\gamma}_{ L, n}^{(1)}(\Delta_{\phi})\,{\textstyle{\frac{\Gamma(4\Delta_{\phi}+4n)\Gamma(2\Delta_{\phi}+p)^2}{\Gamma(2\Delta_{\phi}+2n)^2\Gamma(4\Delta_{\phi}+2n+p)\Gamma(p-2n+1)}}}\,
 =\sum_{k=0}^{2L}(-1)^{k} c_{k,L} P_{\Delta_{\phi}+L}(2\Delta_{\phi}+p+k) \, .
\end{align}
Notice that the sum on the l.h.s. of \eqref{system} is truncated by the $\Gamma(p-2n+1)$ in the denominator. This means that, at fixed value of $p$, equation \eqref{system} provides an invertible linear system which can be solved for $\gamma$. The system can also be inverted explicitly using the identity
\begin{align}
\sum_{p=0}^{2n} \frac{(4 m+4\Delta_{\phi} -1)\, \Gamma(p+4\Delta_{\phi} +2m-1) (-1)^{p}}{\Gamma(2m-p+1)\,\Gamma(2\Delta+p)^2} \frac{\Gamma(2\Delta_{\phi}+p)^2}{\Gamma(4\Delta_{\phi}+2n+p)\Gamma(p-2n+1)}=\d_{m, n}
\end{align}
yielding \eqref{greatresult}, which we rewrite here
\begin{align}
 \hat{\gamma}_{ L, n}^{(1)}(\Delta_{\phi})&=\frac{\Gamma(L+\Delta_{\phi})^4}{\Gamma(2L+2\Delta_{\phi})}\sum_{p=0}^{2n} \sum_{k=2L-p}^{2L}\sum_{l=0}^{k+p-2L}(-1)^k c_{k,L} \times\label{gammahatsums}\\   &\frac{(4\Delta_{\phi}+2n-1)_p (-2n)_p (2L-k-p)_l (1-\Delta_{\phi}-L)_l (2\Delta_{\phi}+k+p)_l (\tfrac12)_l}{(l!)^3 (2\Delta_{\phi})_p (\tfrac12+\Delta+L)_l} \nonumber  \,.
\end{align}
Equation \eqref{gammahatsums} has been derived under the assumption that $\Delta_{\phi}$ takes integer values. However one can argue that it holds for any $\Delta_{\phi}$ by noticing that the result for $L=0$ agrees with that of \cite{Mazac:2018ycv}, which has been obtained without assuming integer $\Delta_{\phi}$. Furthermore, equation \eqref{ML} implies that the CFT data at $L=0$ fully determine those at higher $L$ or, in other words, equation \eqref{gammahatsums} could be rewritten as a combination of anomalous dimension for $L=0$. This is enough to show that \eqref{gammahatsums} holds for any $\Delta_{\phi}$.

Notice that all the sums in \eqref{gammahatsums} are finite, so for a given value of $L$ and $n$ it is straightforward to extract the value of the anomalous dimension $\hat{\gamma}_{ L, n}^{(1)}(\Delta_{\phi})$. It turns out expression \eqref{gammahatsums} can be rewritten as
\begin{align}\label{gammaLN2}
  \hat{\gamma}_{ L, n}^{(1)}(\Delta_{\phi})&=\hat{\mathcal{G}}_{L,n}(\Delta_{\phi}) \hat{\mathcal{P}}_{L,n}(\Delta_{\phi})\,,
\end{align}
where
\begin{align}\label{gammaLN3}
 \hat{\mathcal{G}}_{L,n}(\Delta_{\phi})=\tfrac{\sqrt{\pi }  4^{-2 \Delta -L+1} \Gamma (2 \Delta )^2\Gamma (L+\frac{1}{2}) \Gamma (L+\Delta )^4 \Gamma (L+2 \Delta -\frac{1}{2}) \Gamma
   (n+\Delta +\frac{1}{2}) \Gamma (L-n+\Delta )}{\Gamma (L+1) \Gamma (L+\Delta +\frac{1}{2})^2 \Gamma (L+2 \Delta ) \Gamma (n+\Delta )^3 \Gamma
   (2 n+2 \Delta -\frac{1}{2}) \Gamma (L+n+\Delta +\frac{1}{2})} \,,
\end{align}
while $\hat{\mathcal{P}}_{L,n}(\Delta_\phi)$ is a polynomial in $n$ and in $\Delta_{\phi}$ of degree $6L$. It is easy to extract these polynomials from \eqref{greatresult}, but since they are very long we attach to the paper a short \texttt{Mathematica} notebook where the function $\texttt{FindPolynomial[L,$\Delta$,n]}$ allows to extract $\hat{\mathcal{P}}_{L,n}$ for many values of $L$ (the function gets slower and slower at higher $L$, but in principle it works for any $L$).

To make contact with the bootstrap results, the coefficients $a_\ell$  in the definition~\eqref{ML} of  $M_L(s)$ should be chosen so as to have the  bootstrap normalization, namely $\gamma^{(1)}_{\ell,L}=0$ for $0\leq \ell< L$, and $\gamma^ {(1)}_{L,L} =1$.  In this case we have
\begin{align}\label{gammaLB2}
	\gamma_{ L, n}^{(1)}(\Delta_{\phi})&=\mathcal{G}_{L,n}(\Delta_{\phi}) \mathcal{P}_{L,n}(\Delta_{\phi}) \,,
\end{align}
where
\begin{align}\label{gammaLB3}
	\mathcal{G}_{L,n}(\Delta_{\phi})=\frac{4^{-L} \left(L+\frac{1}{2}\right)_{\Delta _{\phi }} \left(L+\Delta _{\phi }\right)_{\Delta _{\phi }} (-L+n+1)_{\Delta _{\phi }-1} \left(L+n+\Delta _{\phi }+\frac{1}{2}\right)_{\Delta _{\phi }-1}}{\Gamma \left(\Delta _{\phi }\right) \left(\Delta _{\phi }\right)_{3 L} \left(2 L+\Delta _{\phi }+\frac{1}{2}\right)_{\Delta _{\phi }-1} \left(L+2 \Delta _{\phi }-\frac{1}{2}\right)_{2 L} \left(n+\frac{1}{2}\right)_{\Delta _{\phi }} \left(n+\Delta _{\phi }\right)_{\Delta _{\phi }}} \,,
\end{align}
while $\mathcal{P}_{L,n}(\Delta_\phi)$ is a polynomial of degree $4L$ in $n$ and $5L$ in $\Delta_\phi$. The explicit polynomials for the first few values of $L$ are detailed in Appendix \ref{Ap: anomalous dimension} and, up to $L=3$, they perfectly agree with the result of \cite{Ferrero:2019luz}. The attached \texttt{Mathematica} notebook has values of $L$ ranging from $L=0$ to $L=8$ as well as a function~\texttt{FindBootstrapPolynomial[L,$\Delta$,n]} to compute $\mathcal{P}_{L,n}(\Delta_\phi)$ for arbitrary $L$.

\subsection{Alternative formulation of Mellin amplitude}  
\label{sec:alternativeMellin}
 
We conclude this section by pointing out that there is another noteworthy definition of Mellin transform,  identified by taking $a=-2\Delta_{\phi}+1$ in equation (\ref{Mellin1param}):
\be\label{Mellin2}
\mathcal{M}_{-2\Delta_{\phi}+1}(s)\equiv\tilde M_{\Delta_{\phi}}(s)=\int_0^\infty dt\, f(t)\,\left(\frac{t}{1+t}\right)^{-2\Delta_{\phi}+1}\,t^{-1-s} .
\ee 
The interesting feature of this particular choice is to provide the simplest representation for Mellin amplitudes of $\bar D$-functions.  We can motivate this claim by looking at the more general expression (\ref{Mellin1param}), where $a$ is a free parameter. We rewrite this in terms of the $\bar D$-functions using the identity (\ref{gandf})
\begin{equation} \label{altMellina}
\begin{aligned}
\mathcal{M}_a(s) =\int_{0}^{\infty} d t\, \bar{D}_{\Delta_{\phi}\Delta_{\phi}\Delta_{\phi}\Delta_{\phi}}(t) \left(\frac{t}{1+t}\right)^{2\Delta_{\phi}}\left(\frac{t}{1+t}\right)^{a} t^{-1-s}
\end{aligned}.
\end{equation}
In particular, if we consider for example $\bar D_{1111}$, we obtain
\begin{equation}\label{resaltMellina}
\begin{aligned}
\mathcal{M}_a(s) =\tfrac{2 \Gamma(s-1) \Gamma(a+2\Delta_{\phi}-s)(\psi(s-1)-\psi(a+2\Delta_{\phi}-s))}{\Gamma(a+2\Delta_{\phi}-1)} +\tfrac{2 \Gamma(s-1) \Gamma(a+2\Delta_{\phi}-s-1)(\psi(a+2\Delta_{\phi}-2)-\psi(s-1))}{\Gamma(a+2\Delta_{\phi}-2)}
\end{aligned}.
\end{equation}
This expression simplifies for an integer value of the parameter $a$ below a certain threshold, namely $a \leq -2\Delta_{\phi}+1$. Considering the region of convergence of (\ref{altMellina}), we need 
\begin{equation} 
\begin{aligned}
\Delta_0 > \Delta_{\phi}-\frac{a}{2} ,
\end{aligned}
\end{equation}
and given that $\Delta_0=2\Delta_{\phi}$,  the only simple convergent integral has $a=2\Delta_{\phi}+1$,  yielding (\ref{Mellin2}) and the simple Mellin amplitude for $\bar{D}_{1111}$
\begin{equation} 
\begin{aligned}
\tilde M_{1}(s)= 2\, \Gamma(-s)\Gamma(s-1).
\end{aligned}
\end{equation}
See equations (\ref{MellinL0}) and (\ref{altMellinpat}) to appreciate the difference between the two representations of the $\bar{D}$-functions, obtained respectively with (\ref{MellinDbar}) and (\ref{Mellin2}).\\
Apart from this property,  (\ref{Mellin2}) satisfies \eqref{crossingmellin}, which reads
\be\label{crossing2gen}
\tilde{M}(s)=\tilde{M}(1-s)\,.
\ee
Using (\ref{blockexp}) and (\ref{cross-f-t}), we can then derive the strip of convergence of this Mellin definition
\be\label{convergence2gen}
2\Delta_\phi-\Delta_0<\text{Re}(s)<1+\Delta_0-2\Delta_\phi\,,
\ee
which translates, perturbatively ($\Delta_0=2\Delta_\phi$), in 
\be\label{convergence2}
0<\text{Re}(s)<1~.
\ee
The inverse Mellin transform reads 
\be
f(t)=\int_{c-i\infty}^{c+i\infty}\frac{ds}{2\pi i}\,\tilde{M}(s)\left(\frac{t}{1+t}\right)^{2\Delta_{\phi}-1}\,t^s
\ee
where the range of the real constant $c$ is the same of $\text{Re}(s)$ in \eqref{convergence2gen}, and therefore in perturbation theory the contour of the integral in the complex $s$-plane is any straight line within the interval \eqref{convergence2}. \\
We can finally report a general structure for the Mellin transform of the $\bar{D}$-functions
\be \label{altMellinpat}
\tilde M_{\Delta_\phi}(s) = P_{\Delta_{\phi}} (s)\, \Gamma(-s-2\Delta_{\phi}+2) \Gamma(s-2\Delta_{\phi}+1)\,,
\ee
where 
\begin{equation}
\resizebox{.97\hsize}{!}{$P_{\Delta_{\phi}} (s) = 2\, \Gamma(2\Delta_{\phi}-1) {}_{4}F_{3} \left(\{-s-2\Delta_{\phi}+2,s-2\Delta_{\phi}+1,1-\Delta_{\phi},1-\Delta_{\phi}\};\{2-2\Delta_{\phi},2-2\Delta_{\phi},2-2\Delta_{\phi}\};1\right)$}.
\end{equation}
Note that $P_{\Delta_{\phi}}$ is a polynomial for integer $\Delta_{\phi}$, that we now tabulate for the first few cases, using a more convenient rewriting, $Q_{\Delta_{\phi}} (s(s-1))= P_{\Delta_{\phi}}(s)$
\begin{align} \label{DfunctMellintable2}
{\renewcommand\arraystretch{1.3} 
\begin{tabular}{ c | c }
$\Delta_{\phi}$ & Q$_{\Delta_{\phi}} (x)$ \\
\hline 
1  &  2 		\\																										
2  & $2\,(2+x)	$ 	\\																									
3  & $32 \,(24 + 22 \,x + x^2)$		\\											
4  & $2592\,(720 + 876  \,x + 100 \, x^2 + x^3)$	\\
5  & $663552 \,(40320 + 58416  \,x + 10508 \, x^2 + 300 \, x^3 + x^4)$	\\
\end{tabular}      }   
\end{align}
To conclude, we report an alternative closed-form expression valid for integer value of $\Delta_{\phi}$
\be
\tilde{M}_{\Delta_\phi}(s)=\! \sum^{\Delta_\phi-1}_{n=0}\frac{2\,(-1)^n\,\Gamma(\Delta_\phi)^2\Gamma(2\Delta_\phi-1-n)^3}{\Gamma(n+1)\Gamma(\Delta_\phi-n)^2}  
\Gamma(-s-2\Delta_{\phi}+2+n)\, \Gamma(s-2\Delta_\phi+1+n)\,,
\ee
which is a linear combination of squared Gamma functions. \\
Despite this nice representation of the $\bar{D}$-functions, the correspondence between the poles and the physical exchanged operators is more obscure, in contrast with (\ref{leftpoles}) and (\ref{rightpoles}) for the Mellin transform (\ref{Mellin}).  We therefore reckoned that the Mellin transform defined in (\ref{Mellin}) is the most suitable for the applications we presented in this paper, which have as a main goal the extraction of CFT data.

\section*{Acknowledgements}

We thank Pietro Ferrero for very useful discussions.  The research of LB is funded through the MIUR program for young researchers ``Rita Levi Montalcini''. The research of GB and VF is funded from the European Union's Horizon 2020 research and innovation programme under the Marie Sklodowska-Curie ITN grant No 813942.   The research of VF received also funding from the STFC grant ST/S005803/1, from the STFC Consolidated Grant
Theoretical Physics at City University ST/P000797/1, and from the Einstein Foundation Berlin through an Einstein Junior Fellowship.  The research of GP is funded from the Einstein Foundation Berlin and by the Deutsche Forschungsgemeinschaft (DFG, German Research Foundation) - Projektnummer 417533893/GRK2575 ``Rethinking Quantum Field Theory'', from which also GB and VF benefit.

 
 \appendix

\section{Poles and series}
\label{app:polesandseries}
In this appendix we address some subtleties in the perturbative expansion of the Mellin amplitudes in the context of the analytic sum rules. In all the following, we will consider a function $\hat{F}(s)$ defined nonperturbatively and its expansion in a small parameter $\lambda$, as well as the functional 
\begin{align}
	\omega_K[\hat{F}] = \oint_{\mathbb{C}|_{\infty}} \frac{ds}{2 \pi i} \hat{F}(s) K(s),
\end{align}
defined as the contour integral of $\hat{F}(s)$ over the circle at $\infty$ in the complex plane with an integration kernel $K(s)$. This functional is well defined and vanishing for meromorphic functions $\hat{F}(s)$ and $K(s)$ such that
\begin{align}\label{convergencecondition}
	\hat{F}(s)K(s)\underset{|s|\rightarrow\infty}{\sim} s^{-1-\alpha} \qquad \alpha>0.
\end{align}
\subsection{Nonperturbative zeros and perturbative poles}
The nonperturbative Mellin function $\hat{M}(s)$ has zeros at positions $s=2\Delta+n$, while the perturbative expansion has poles in those positions. The reason for this and the subtleties of evaluating the sum rule will be illustrated with an example.\par Let $f(s)$ be a well-behaved function\footnote{By this, we mean that $f$ is meromorphic, has no essential singularities and behaves as $s^{-\alpha}$ at large s, where $\alpha$ is a positive parameter. For simplicity, we take $f(s)$ to be regular, so without poles. However, including these poles (different to those of $\hat{F}$) is  straightforward and does not change the conclusions.}. We define the $\lambda$-dependent function
\begin{align}
	\hat{F}(s) = \frac{(s-1)f(s)}{s-1-\lambda}.
\end{align}
Upon expanding $F(s)$ in the parameter $\lambda$, we obtain the geometric series
\begin{align}
	\hat{F}(s) =\sum_{k=0}^\infty \left(\frac{\lambda}{s-1}\right)^k f(s),
\end{align}
whose radius of convergence is $\lambda<|s-1|$.
This perturbative series cannot be evaluated at finite $\lambda$ in $s=1$, however several features are noteworthy. At $s=1$, we have poles of increasing degree at each order in the $\lambda$ expansion while the nonperturbative expression has a simple zero at that point. Let us evaluate the contour integral
\begin{align}\label{Eq: Functional 1}
	\omega_1[\hat{F}] = \oint \frac{ds}{2\pi i} \frac{\hat{F}(s)}{s-1}
\end{align}
in three different ways. Firstly, by inserting the nonperturbative $\hat{F}$
\be
	\oint \frac{ds}{2\pi i} \frac{\hat{F}(s)}{s-1} = 	\oint \frac{ds}{2\pi i} \frac{f(s)}{s-1-\lambda}
	=f(1+\lambda).
\ee
This method is the most intuitive way of evaluating (\ref{Eq: Functional 1}) and is perfectly well defined. However, it requires the knowledge of the full nonperturbative function, which is generally unknown. \par Secondly, we can consider separately the poles of the integration kernel $\frac{1}{s-1}$ from those of the function $\hat{F}$:
\begin{align}
	\oint \frac{ds}{2\pi i} \frac{\hat{F}(s)}{s-1} &=\textrm{Res}(\frac{1}{s-1}) \hat{F}(1)+\sum_{s^*}\frac{\textrm{Res}(\hat{F})(s^*)}{s^*-1}=f(1+\lambda).
\end{align}
This is not identical to the previous method, since it assumes more (notably that the poles of $K(s)$ and those of $\hat{F}$, $s^*$ above, are distinct) and requires less information from the nonperturbative function; only the residues of $\hat{F}(s)$ and the value of $\hat{F}(s_i)$ at a finite number of points are required to evaluate (\ref{Eq: Functional 1}).\par  Finally, one can use the series expansion of $\hat{F}(s)$
\begin{align}
	\omega_1[\hat{F}]&=\oint \frac{ds}{2\pi i}\left(\sum_{k=0}^\infty \frac{\lambda^k}{(s-1)^{k+1}} f(s)\right) =\sum_{k=0}^{\infty}\frac{\lambda^k f^{(k)}(1)}{k!},
\end{align}
which gives the $\lambda$-expansion of $f(1+\lambda)$. Under the assumption that $f$ is analytic near~1, we can then resum the Taylor series to obtain $f(1+\lambda)$ and obtain the same result as with the previous two methods. In this specific example, the only operation which is not allowed is to evaluate $\hat{F}(1)$ using the nonperturbative series, since the latter has a vanishing radius of convergence for $s=1$. In the next subsection we will explore a setting where the perturbative evaluation of the functional is problematic and requires the truncation of the series at a fixed order.\par 
\subsection{Bad Regge behaviour}
We now consider another pathological case where the perturbative expansion of the correlator makes the large $s$  behaviour worse than that of the nonperturbative expression. To illustrate this we take the example
\begin{align}\label{Fbad}
	\hat{F}(s) &= \frac{f(s)}{1-\lambda s^2},
\end{align}
and evaluate the functional
\begin{align}\label{Eq: Functional 2}
	\omega_3[\hat{F}] &= \oint_{\mathbb{C}|_{\infty}}\frac{ds}{2\pi i}\frac{\hat{F}(s)}{s(s^2-1)}.
\end{align}
The convergence of the integral~(\ref{Eq: Functional 2}) imposes a bound on the large $s$ behaviour of $f(s)$ in~\eqref{Fbad}\footnote{As in the previous case, we consider that $f(s)$ is meromorphic in $s$ and ignore its poles, since they act as spectators in the comparison between the perturbative and nonperturbative case. }
\begin{align}
	f(s) &\underset{s\rightarrow\infty}{\sim}s^{4-\alpha}& \alpha &>0.
\end{align}
The functional (\ref{Eq: Functional 2}) can then be evaluated explicitly as
\begin{align}\label{Eq: functional non pert}
	\omega_3[\hat{F}]  =-f(0)+  \frac{f(1)+f(-1)}{2(1-\lambda) }-\frac{\lambda }{2(1-\lambda) } \left(f\left(\frac{1}{\sqrt{\lambda }}\right)+f\left(-\frac{1}{\sqrt{\lambda }}\right)\right).
\end{align}
The perturbative expansion of $\hat{F}(s)$ is 
\begin{align}
	\hat{F}(s)=f(s)+\lambda s^2 f(s)+\lambda^2 s^4 f(s)+O(\lambda^3).
\end{align}
In this perturbative setting, even with a stricter condition on the large $s$ behaviour of $f(s)$
\begin{align}\label{Eq: convergence Int}
	f(s)&\underset{s\rightarrow\infty}{\sim} s^{-\alpha}&\alpha&>0,
\end{align} 
we must truncate the series at order $\lambda$ in order to evaluate the functional 
\begin{align}\label{Eq functional Order lambda}
	\omega_3[f(s)+\lambda s^2 f(s)] &= -f(0)+\frac{f(1) +f(-1)}{2} + \lambda \frac{f(1)+f(-1)}{2}.
\end{align}
Comparing (\ref{Eq functional Order lambda}) to the small-$\lambda$ expansion of (\ref{Eq: functional non pert})
\begin{align}
	\omega_3[\hat{F}]&=-f(0)+\frac{f(1) +f(-1)}{2} + \lambda \frac{f(1)+f(-1)}{2}-\frac{\lambda }{2}(f(\frac{1}{\sqrt{\lambda }})+f(-\frac{1}{\sqrt{\lambda }})) +o(\lambda),
\end{align}
we see that the condition of convergence (\ref{Eq: convergence Int}) is exactly what is needed to get rid of the final terms and find agreement between the results
\begin{align}
	-\frac{\lambda }{2}(f(\frac{1}{\sqrt{\lambda }})+f(-\frac{1}{\sqrt{\lambda }}))& \underset{\lambda \rightarrow0}{\sim}\lambda^{1+\frac{\alpha}{2}}\underset{\lambda \rightarrow0}{=}  o(\lambda).
\end{align}
We therefore have agreement between the functional of the truncated $\lambda$-expansion of $\hat{F}(s)$ and the truncated $\lambda$-expansion of the functional of $\hat{F}(s)$
\begin{align}
	\omega_3[\hat{F}|_{\lambda}]  = \left(	\omega_3 [\hat{F}]\right)|_{\lambda},
\end{align}
where the order at which we are required to truncate is controlled by the large $s$ behaviour of $K$ (which can always be chosen to satisfy the convergence condition~\eqref{convergencecondition} at a given order of expansion).

 \section{Sum rules for other generalized free field theories}\label{otherGFF}
 
 In this appendix we apply the sum rules discussed in Section~\ref{sumrules} to  other examples of generalised free theories (GFF) in which the spectrum of exchanged operators is known. For these theories, sum rules are obtained inserting~\eqref{GFFkernel} into~\eqref{functional} and using the GFF spectrum, and read
 \begin{align}
 \!\!\!\! \sum_{\D,k} c_\D \frac{(-1)^{k+1}\Gamma(2\D)\Gamma(\D+k)}{\Gamma(\D)^2\Gamma(2\D+k)\Gamma(2\Delta_\phi-\D-k)\Gamma(k+1)}(\frac{1}{\Delta+k+p}-\frac{1}{2\Delta_\phi-\Delta-k+p})=0\,.
 \end{align}
 The GFF spectrum $\Delta = 2\Delta_\phi+n$ for exchanged operators has the effect of truncating the sum above, because of the factor of $\Gamma(2\Delta_\phi-\Delta-k)$ in the denominator. The example of bosonic GFF in Section \ref{subsubsec:GFFsumrules} illustrates the case of even $n$, namely $\Delta=2\Delta_\phi+2n$.  Here we extend this to  odd integers, namely  $\Delta=2\Delta_\phi+2n+1$,  in covering the free fermionic model. We also consider the free bosonic and fermionic models with O(N) symmetry.

\bigskip

 In the case of a free fermion theory, the spectrum $\Delta=2\Delta_\phi+2n+1$ of exchanged operators leads to the sum rule
 \begin{align} \label{Sum rule GFF fermions}
 	\sum_{n=0}^p c_{n}\frac{2 \Gamma (2 p+1) \Gamma (2 (p+\Delta_\phi )) \left(\Delta_\phi -2 \Delta_\phi  p+2 n^2+4 \Delta_\phi  n+n\right) \Gamma (4 n+4 \Delta_\phi +2)}{\Gamma (2 p-2 n+1) \Gamma (2 n+2 \Delta_\phi +1)^2 \Gamma (2 (p+n+2 \Delta_\phi +1))}=0.
 \end{align}
Just like in the bosonic case, see~\eqref{bosonicGFFsumrule}, this is a recursive relation for the OPE coefficients, whose general solution is
 \begin{align}\label{OPE fermion}
 	c_n^{(0)}=\frac{2 \Gamma (2 n+2 \Delta_\phi +1)^2 \Gamma (2 n+4 \Delta_\phi )}{\Gamma (2 \Delta_\phi ) \Gamma (2 \Delta_\phi +1) \Gamma (2 n+2) \Gamma (4 n+4 \Delta_\phi +1)}\,.
 \end{align}
 This is confirmed by the vanishing of \eqref{Sum rule GFF fermions}. As usual, this is true up to an overall scaling, and the choice in \eqref{OPE fermion} is set by requiring $c_0^{(0)}=1$.
 
\bigskip

Let us consider a bosonic four-point function with $O(N)$ symmetry. We write the Mellin amplitude as  a sum of the singlet, antisymmetric and traceless symmetric contributions
 \begin{align}
 	\hat{\mathcal{M}}^{1234}(s)  = 	\hat{\mathcal{M}}^S\delta^{12}\delta^{34}+\hat{\mathcal{M}}^A(\delta^{13}\delta^{24}-\delta^{14}\delta^{23})+\hat{\mathcal{M}}^T(\frac{\delta^{13}\delta^{24}+\delta^{14}\delta^{23}}{2}-\frac{\delta^{12}\delta^{34}}{N}),
 \end{align}
with scalar coefficient functions $\hat{\mathcal{M}}^{S}(s)$, $\hat{\mathcal{M}}^{A}(s)$ and $\hat{\mathcal{M}}^{T}(s)$.  
   In these channels, exchanged operators will be of the form $\phi_i\partial^{2n}_x\phi^i$,$\phi^{[i}\partial^{2n+1}_x\phi^{j]}$,$\phi^{(i}\partial^{2n}_x\phi^{j)}$ respectively, with same spectra of exchanged operators previously seen ($\Delta=2\Delta_\phi+2n$,  $\Delta=2\Delta_\phi+2n+1$,   $\Delta=2\Delta_\phi+2n$ respectively).   Therefore we get the corresponding OPE coefficients 
 \begin{align}
 	c_n^{S} & = 	c^S_0 \frac{2 \Gamma (2 n+2 \Delta_\phi )^2 \Gamma (2 n+4 \Delta_\phi -1)}{\Gamma (2 \Delta_\phi )^2 \Gamma (2 n+1) \Gamma (4 n+4 \Delta_\phi -1)}\,,\\
 	 	c_n^{A} & = 	c^A_0 \frac{2 \Gamma (2 n+2 \Delta_\phi +1)^2 \Gamma (2 n+4 \Delta_\phi )}{\Gamma (2 \Delta_\phi )\Gamma (2 \Delta_\phi+1 )\Gamma (2 n+2) \Gamma (4 n+4 \Delta_\phi +1)}\,,\\
 c_n^{T} & = 	c^T_0 \frac{2 \Gamma (2 n+2 \Delta_\phi )^2 \Gamma (2 n+4 \Delta_\phi -1)}{\Gamma (2 \Delta_\phi )^2 \Gamma (2 n+1) \Gamma (4 n+4 \Delta_\phi -1)}\,,
 \end{align}
up to a normalisation factor which can be easily found by looking at the first identity contribution in the different channels. This gives the known result (see for example~\cite{Ferrero:2019luz})
 \begin{align}
 	c_n^{S} & = 	 \frac{1}{N} \frac{2 \Gamma (2 n+2 \Delta_\phi )^2 \Gamma (2 n+4 \Delta_\phi -1)}{\Gamma (2 \Delta_\phi )^2 \Gamma (2 n+1) \Gamma (4 n+4 \Delta_\phi -1)}\,,\\
 	c_n^{T} & = 	 \frac{2 \Gamma (2 n+2 \Delta_\phi )^2 \Gamma (2 n+4 \Delta_\phi -1)}{\Gamma (2 \Delta_\phi )^2 \Gamma (2 n+1) \Gamma (4 n+4 \Delta_\phi -1)}\,,\\ 
 	c_n^{A} & = 	- \frac{2 \Gamma (2 n+2 \Delta_\phi +1)^2 \Gamma (2 n+4 \Delta_\phi )}{\Gamma (2 \Delta_\phi )^2\Gamma (2 n+2) \Gamma (4 n+4 \Delta_\phi +1)}\,.
 \end{align}
The same procedure for free fermions with O(N) symmetry leads to  
 \begin{align}
 	c_n^{S} & = \frac{1}{N}	\frac{2 \Gamma (2 n+2 \Delta_\phi +1)^2 \Gamma (2 n+4 \Delta_\phi )}{\Gamma (2 \Delta_\phi ) \Gamma (2 \Delta_\phi +1) \Gamma (2 n+2) \Gamma (4 n+4 \Delta_\phi +1)}\,, \\
 	c_n^{T} & = \frac{2 \Gamma (2 n+2 \Delta_\phi +1)^2 \Gamma (2 n+4 \Delta_\phi )}{\Gamma (2 \Delta_\phi ) \Gamma (2 \Delta_\phi +1) \Gamma (2 n+2) \Gamma (4 n+4 \Delta_\phi +1)} \,, \\ 
 	c_n^{A} & = 	- \frac{2\Delta_\phi \Gamma (2 n+2 \Delta_\phi )^2 \Gamma (2 n+4 \Delta_\phi -1)}{\Gamma (2 \Delta_\phi )^2 \Gamma (2 n+1) \Gamma (4 n+4 \Delta_\phi -1)}\,.
 \end{align}

\section{D-functions}
\label{app:Dfunctions}
The quartic contact diagrams with external conformal dimensions $\D_i$ are expressed in terms of $D$-functions \cite{Liu:1998ty,DHoker:1999kzh, Dolan:2003hv},  defined  for the general case of $AdS_{d+1}$ as
\be \label{D-function}
\!\!\!\!\!\!\!\!
D_{\Delta_{1}\Delta_{2}\Delta_{3}\Delta_{4}}(x_1,x_2,x_3,x_4) =\! \!\int \!\!\frac{dz d^dx}{z^{d+1}} 
\tilde{K}_{\Delta_{1}}\!(z,x;x_1) \tilde{K}_{\Delta_{2}}\!(z,x;x_2) \tilde{K}_{\Delta_{3}}\!(z,x;x_3) \tilde{K}_{\Delta_{4}}\!(z,x;x_4)\,
\ee
via the bulk-to-boundary propagator in $d$ dimensions
\be \label{bulktoboundary}
K_{\Delta}(z,x;x') = {\cal C}_{\Delta} \Big[\frac{z}{z^2+(x-x')^2}\Big]^{\Delta} \equiv  {\cal C}_{\Delta}\,  \tilde{K}_{\Delta}(z,x;x')\,,
\,\qquad {\cal C}_{\Delta_{\phi}} =\frac{\Gamma\left(\Delta_{\phi}\right)}{2\,\sqrt{\pi}\,\Gamma\left(\Delta_{\phi}+{1\ov 2}\right)} \,.
\ee
For vertices with derivatives, the following identity is useful 
\bal 
&g^{\m\n}\partial_\m \tilde{K}_{\Delta_1}(z,x;x_1)\ \partial_\n\tilde{K}_{\Delta_2}(z,x;x_2) 
\\   & \qquad = 
\Delta_1 \Delta_2
\left[\tilde{K}_{\Delta_1}(z,x;x_1)\tilde{K}_{\Delta_2}(z,x;x_2)-2x_{12}^2 \tilde{K}_{\Delta_1+1}(z,x;x_1)\tilde{K}_{\Delta_2+1}(z,x;x_2)\right]\ \,,
\label{identityder}
\eal
where $g^{\m\n}= {z^2}\delta^{\m\n}$ and $\del_\m=(\del_z,\del_r)$,  $r=0,1, 2, ...,d-1$.  

To make explicit the covariant form of the correlator it is useful to introduced the "reduced" functions $\bar D$~\cite{Dolan:2003hv},  defined as  ($\Sigma \equiv \frac{1}{2}\sum_i \Delta_i$) 
\be \label{Dbar}
\!\!\!\!
D_{\Delta_{1}\Delta_{2}\Delta_{3}\Delta_{4}}= 
\frac{\pi^{d\ov 2}\Gamma\left(\Sigma-{d\ov2}\right)}{2\, \Gamma\left(\Delta_1\right)\Gamma\left(\Delta_2\right)\Gamma\left(\Delta_3\right)\Gamma\left(\Delta_4\right)}
\frac{x_{14}^{2(\Sigma-\Delta_1-\Delta_4)} x_{34}^{2(\Sigma-\Delta_3-\Delta_4)}}
{x_{13}^{2(\Sigma-\Delta_4)} x_{24}^{2\Delta_2}}\bar{D}_{\Delta_1\Delta_2\Delta_3\Delta_4}(u,v)
\ee
and depending only on the cross-ratios $u=\frac{x_{12}x_{34}}{x_{13}x_{24}}\,,  v=\frac{x_{14}x_{23}}{x_{13}x_{24}}$.  Their explicit expression in terms a Feynman 
parameter integral reads in the general case
\begin{equation}
	\bar{D}_{\Delta_1\Delta_2\Delta_3\Delta_4}(u,v)=
	\int d\alpha d\beta d\gamma\  \delta(\alpha+\beta+\gamma-1)\ 
	\alpha ^{\Delta _1-1} \beta ^{\Delta _2-1} \gamma ^{\Delta _3-1} 
	\frac{\Gamma \left(\Sigma-\Delta _4\right) \Gamma\left(\Delta_4\right)}
	{\big(\alpha  \gamma + \alpha  \beta\, u  + \beta  \gamma\, v\big)^{\Sigma-\Delta_4}}\,.
	\label{Dbar-integral}
\end{equation}
In  $d=1$ as usual they only depend on the single variable   $z$ ($u=z^2$, $v=(1-z)^2$),
\be\label{Dbar1d}
\bar D_{\Delta\Delta\Delta\Delta}(z)=\frac{\Gamma(\Delta)^4}{\Gamma(2\Delta)} (1-z)^{-2\Delta}\!\int_{-\infty}^{+\infty}\!d\tau\,e^{-\tau} {}_2F_1\big(\Delta,\Delta,2\Delta,\textstyle-\frac{4z}{(1-z^2)}\cosh^2\frac{\tau}{2}\big)\, .
\ee

Some explicit expression for $\bar{D}$-functions  read
\begin{eqnarray}\label{Dbar-explicit}
	\bar D_{1111}&=&-\frac{2 \log (1-z )}{z }-\frac{2 \log (z )}{1-z }\, ,\\
	\bar{D}_{2222} &=&-\frac{2 \left(z ^2-z +1\right)}{15 (1-z )^2 z ^2}+\frac{\left(2 z ^2-5 z +5\right) \log (z )}{15 (z -1)^3}-\frac{\left(2 z ^2+z +2\right) \log (1-z )}{15 z ^3}\, ,\\
	\bar{D}_{3333}&=&\frac{\left(8 z ^4-36 z ^3+64 z ^2-56 z +28\right) \log(z)}{105 (z -1)^5}+\frac{\left(-8 z ^4-4 z ^3-4 z ^2-4 z -8\right) \log (1-z )}{105 z ^5}\nonumber\\
	&&+\frac{-24 z ^6+72 z ^5-74 z ^4+28 z ^3-74 z ^2+72 z -24}{315 (z -1)^4 z ^4}\, .\label{Dbar-explicit-end}
\end{eqnarray}

Further expressions can be found through the identities in~\cite{Dolan:2003hv}.  Useful relations between $\bar D$-function of consequent weight are
%
\begin{align}\label{Dbarconseq}
	\Delta\,\bar{D}_{\Delta \Delta \Delta \Delta }&= \bar{D}_{\Delta \Delta \Delta+1 \Delta+1}+\bar{D}_{\Delta \Delta+1 \Delta \Delta +1}+\bar{D}_{\Delta+1 \Delta \Delta \Delta +1} \, ,\\
\label{Dfunctioncrossing1}
	(\Delta_2+\Delta_4-\Sigma) \,\bar{D}_{\Delta_1 \Delta_2 \Delta_3 \Delta_4} &= \bar{D}_{\Delta_1 \Delta_2+1 \Delta_3 \Delta_4+1} - \bar{D}_{\Delta_1+1 \Delta_2 \Delta_3+1 \Delta_4}\, ,\\	
	(\Delta_1+\Delta_4-\Sigma) \,\bar{D}_{\Delta_1 \Delta_2 \Delta_3 \Delta_4}& =\bar{D}_{\Delta_1+1 \Delta_2 \Delta_3 \Delta_4+1}-(1-z)^2 \bar{D}_{\Delta_1 \Delta_2+1 \Delta_3+1 \Delta_4}\, ,\\
	(\Delta_3+\Delta_4-\Sigma)\, \bar{D}_{\Delta_1 \Delta_2 \Delta_3 \Delta_4}&= \bar{D}_{\Delta_1 \Delta_2 \Delta_3+1 \Delta_4+1}-z^2\bar{D}_{\Delta_1+1 \Delta_2+1 \Delta_3 \Delta_4}\, ,\\
\label{Dfunctioncrossing2}
	\bar{D}_{\Delta_1 \Delta_2 \Delta_3 \Delta_4}&=(1-z)^{2(\Delta_1+\Delta_4-\Sigma)}\bar{D}_{\Delta_2 \Delta_1 \Delta_4 \Delta_3}\, ,\\
	&=\bar{D}_{\Sigma-\Delta_3 \Sigma- \Delta_4 \Sigma-\Delta_1 \Sigma-\Delta_2}\, ,\\
	&=z^{2(\Delta_3+\Delta_4-\Sigma)}\bar{D}_{\Delta_4 \Delta_3 \Delta_2 \Delta_1}\, .
\end{align}

 \section{Details on diagrammatic checks}\label{WittenL2}
 In this section we give some details on the diagrammatics checks performed in Section~\ref{diagrammaticchecks}. We compute the four-point function resulting from the $L=2$ interaction term 
 \begin{align}
 \mathcal{L}_{\textrm{int}} =	-g(\partial_\mu \partial _\nu \phi \partial^\mu \partial^\nu \phi)^2,
 \end{align}
to illustrate the convenience of the basis of interaction terms~(\ref{interactionlag}) 
described in the main text and  write the correlator in term of  $f_{L}^{(1)}(z)$  defined in (\ref{ansatzcorrelator}). This interaction leads to the connected part of the four-point correlator
  \begin{align}
 	\langle \phi(x_1)\phi(x_2)\phi(x_3)\phi(x_4)\rangle|_{\textrm{conn}}  = \sum_{\sigma(\{x_1,x_2,x_3,x_4\})} I(x_1,x_2,x_3,x_4),
 \end{align}
 where the integral 
  \begin{align}
 	I(x_1,x_2,x_3,x_4) &=g\int\frac{dx dy}{y^2}(A K_{\Delta_\phi}^4-4 B x_{34}^2K_{\Delta_\phi}^2 K_{\Delta_\phi+1}^2 +4C x_{12}^2x_{34}^2K_{\Delta_\phi+1}^4 \nonumber \\
	\label{IL2}
 	&+8 Dx_{34}^4 K_{\Delta_\phi}^2K_{\Delta_\phi+2}^2-8 Ex_{12}^2 x_{34}^4 K_{\Delta_\phi+1}^2K_{\Delta_\phi+2}^2+16F x_{12}^4x_{34}^4 K_{\Delta_\phi+2}^4).
 \end{align} 
corresponds to one specific choice of Wick contractions. Above, we used the identity (\ref{identityder}) to rewrite the derivatives of propagators, and used the shorthand notation
\begin{align}
	K_{\Delta_\phi+1}^2K_{\Delta_\phi+2}^2 = K_{\Delta_\phi+1}(x,y;x_1)K_{\Delta_\phi+1}(x,y;x_2)K_{\Delta_\phi+2}(x,y;x_3)K_{\Delta_\phi+2}(x,y;x_4).
\end{align}
The constants in \eqref{IL2} are given explicitly by
 \begin{align}
 	A&=\Delta_\phi^8\,,&B&=(\Delta_\phi^4+\Delta_\phi^2(\Delta_\phi+1)^2)\Delta_\phi^4\,,&C&=(\Delta_\phi^4+\Delta_\phi^2(\Delta_\phi+1)^2)^2\nonumber \,,\\
 	D &= \Delta_\phi^6(\Delta_\phi+1)^2\,,&E&=\Delta_\phi^2(\Delta_\phi+1)^2(\Delta_\phi^4+\Delta_\phi^2(\Delta_\phi+1)^2)\,,&F&=\Delta_\phi^4(1+\Delta_\phi^4).
 \end{align}
 Performing the integrals and permutations, we find from the first three terms in~\eqref{IL2} the contribution
 \begin{align}
& 	\sum_{\sigma(\{x_1,x_2,x_3,x_4\})}\int\frac{dx dy}{y^2} A K_{\Delta_\phi}^4= 4! gA k  \bar{D}_{\Delta_\phi \Delta_\phi \Delta_\phi \Delta_\phi}\,,\\
 	&-4 B\sum_{\sigma(\{x_{i=1...4}\})}\int\frac{dx dy}{y^2}   x_{34}^2K_{\Delta_\phi}^2 K_{\Delta_\phi+1}^2=-16kB \frac{(2\Delta_\phi-\frac{1}{2})}{\Delta_\phi} \bar{D}_{\Delta_\phi\Delta_\phi\Delta_\phi\Delta_\phi},\\
 	&4C \!\!\!\sum_{\sigma(\{...\})}\!\!\int\frac{dx dy}{y^2}   x_{12}^2x_{34}^2K_{\Delta_\phi+1}^4 \!= \!32 k C\textstyle{\frac{(2\Delta_\phi+\frac{1}{2})(2\Delta_\phi-\frac{1}{2})}{\Delta_\phi^4} (1+z^2+(1-z)^2)}\bar{D}_{\Delta_\phi+1\Delta_\phi+1\Delta_\phi+1\Delta_\phi+1},
 \end{align}
 where we have use the identity (\ref{Dbarconseq}) from \cite{Dolan:2000ut}.
The next two terms give
 \begin{align}
 &	8 D\sum_{\sigma(\{...\})}\int\frac{dx dy}{y^2}  x_{ij}^4 K_{\Delta_\phi}^2K_{\Delta_\phi+2}^2= 64 k D\frac{(2\Delta_\phi+\frac{1}{2})(2\Delta_\phi-\frac{1}{2})}{\Delta_\phi^2(\Delta_\phi+1)^2} \nonumber \\
 &\qquad\qquad	\times (2\bar{D}_{\Delta_\phi\Delta_\phi\Delta_\phi\Delta_\phi} +(1+z^2+(1-z^2))\bar{D}_{\Delta_\phi+1\Delta_\phi+1\Delta_\phi+1\Delta_\phi+1})
 \end{align}
and
 \begin{align}
 	&-8 E\sum_{\sigma(\{...\})}\int\frac{dx dy}{y^2}x_{12}^2 x_{34}^4 K_{\Delta_\phi+1}^2K_{\Delta_\phi+2}^2=-64 k E\frac{\Gamma(2\Delta_\phi-\frac{1}{2}+3)}{\Gamma(2\Delta_\phi-\frac{1}{2})\Delta_\phi^4(\Delta_\phi+1)^2}   \nonumber \\
 	& \qquad\times\textstyle (\frac{\Delta_\phi^4}{2(3+4\Delta_\phi)}\bar{D}_{\Delta_\phi\Delta_\phi\Delta_\phi\Delta_\phi}+\frac{3+8\Delta_\phi+6\Delta_\phi^2}{2(3+4\Delta_\phi)}(1+z^2+(1-z)^2)\bar{D}_{\Delta_\phi+1\Delta_\phi+1\Delta_\phi+1\Delta_\phi+1}), 
 \end{align}
  where we have used the identities  (\ref{Dfunctioncrossing1}) and (\ref{Dfunctioncrossing2}) also from \cite{Dolan:2000ut} as well as the new identity~(\ref{Eq: New relation}). Finally the last term in~(\ref{IL2})  gives
 \begin{align}
 	16F \sum_{\sigma(\{...\})}\int&\frac{dx dy}{y^2}x_{12}^4x_{34}^4 K_{\Delta_\phi+2}^4 = 128k F\frac{\Gamma(2\Delta_\phi-\frac{1}{2}+4)}{\Gamma(2\Delta_\phi-\frac{1}{2})\Delta_\phi^4(\Delta_\phi+1)^4}   \nonumber  \\
 	&\qquad \times(1+z^4+(1-z)^4)\bar{D}_{\Delta_\phi+2\Delta_\phi+2\Delta_\phi+2\Delta_\phi+2}.
 \end{align}
In all the previous computations, it was useful to factor out the $\phi^4$ normalisation
\begin{align}
	k&=\frac{\sqrt{\pi}
		\Gamma(2\Delta_\phi-\frac{1}{2})}{2\Gamma(\Delta_\phi)^4(x_{13}x_{24})^{2\Delta_\phi}}.
\end{align}
 Combining these contributions, we obtain
 \begin{align}
 &	\langle \phi(x_1)\phi(x_2)\phi(x_3)\phi(x_4)\rangle|_{\textrm{conn}} = 	\frac{\sqrt{\pi}
 		\Gamma(2\Delta_\phi-\frac{1}{2})}{2\Gamma(\Delta_\phi)^4(x_{13}x_{24})^{2\Delta_\phi}} (a_0 \bar{D}_{\Delta_\phi\Delta_\phi\Delta_\phi\Delta_\phi}\\
 	&+a_1(1+z^2+(1-z)^2)\bar{D}_{\Delta_\phi+1\Delta_\phi+1\Delta_\phi+1\Delta_\phi+1}+a_2(1+z^4+(1-z)^4)\bar{D}_{\Delta_\phi+2\Delta_\phi+2\Delta_\phi+2\Delta_\phi+2})\nonumber 
 \end{align}
 with coefficients
 \begin{align}
 	a_0&=-4 \Delta_\phi ^4 (2 \Delta_\phi  (\Delta_\phi  (\Delta_\phi  (21 \Delta_\phi +22)-55)-2)+7)\,,\\
 	a_1&=4 (4 \Delta_\phi -1) (4 \Delta_\phi +1) (2 \Delta_\phi  (2 \Delta_\phi  (\Delta_\phi  (4 \Delta_\phi +7)+7)+7)+3)\,,\\
 	a_2&=8 (4 \Delta_\phi -1) (4 \Delta_\phi +1) (4 \Delta_\phi +3) (4 \Delta_\phi +5)\,,
 \end{align}
 which gives the expansion in terms of the $f_{L}^{(1)}(z)$ basis stated in equation (\ref{lincombL2}).

 \section{Anomalous dimensions for higher derivative interactions}\label{Ap: anomalous dimension}
In this section, we list the various results for the polynomial part of the anomalous dimension in equation (\ref{gammaLB2}). The attached \texttt{Mathematica} notebook has values of $L$ ranging from $L=0$ to $L=8$ as well as a function~\texttt{FindBootstrapPolynomial[L,$\Delta$,n]} to compute $\mathcal{P}_{L,n}(\Delta_\phi)$ for arbitrary $L$ (the function gets slower and slower at higher $L$, but in principle it works for any $L$).

\scriptsize
\begin{align}
	\mathcal{P}_{0,n}(\Delta)&=1\\
	\hspace{1.5cm}\nonumber \\
		\mathcal{P}_{1,n}(\Delta)&=8 (2 \Delta +1) n^4+8 \left(8 \Delta ^2+2 \Delta -1\right) n^3+2 (2 \Delta -1) (2 \Delta +1) (12 \Delta +1) n^2\nonumber \\
		&+\left(64 \Delta ^4-28 \Delta ^2-2 \Delta +2\right) n+\Delta ^2 \left(16 \Delta ^3-13 \Delta -3\right)\\
		\hspace{1.5cm}\nonumber \\
	\mathcal{P}_{2,n}(\Delta)&= 64 (2 \Delta +3) (2 \Delta +5) n^8+128 (2 \Delta +3) (2 \Delta +5) (4 \Delta -1) n^7\nonumber \\
	&+32 (2 \Delta +3) (2 \Delta +5) \left(56 \Delta ^2-22 \Delta -1\right) n^6\nonumber \\
	&+32 (2 \Delta +3) (2 \Delta +5) (4 \Delta -1) (28 \Delta ^2-5 \Delta -5) n^5\nonumber \\
	&+4 (2 \Delta +3) \left(2240 \Delta ^5+4800 \Delta ^4-2924 \Delta ^3-2156
	\Delta ^2+246 \Delta -415\right) n^4\nonumber \\
	&+8 (2 \Delta +3) (4 \Delta -1) (224 \Delta ^5+576 \Delta ^4-158 \Delta ^3-572 \Delta
	^2-243 \Delta -160) n^3\nonumber \\
	&+4 (2 \Delta -1) (2 \Delta +3) (448 \Delta ^6+1392 \Delta ^5+84 \Delta ^4-2183
	\Delta ^3-2091 \Delta ^2-1134 \Delta -105) n^2\nonumber \\
	&+4 (2 \Delta +3) (4 \Delta -1) (64 \Delta ^7+208 \Delta ^6-36 \Delta ^5-605 \Delta
	^4-554 \Delta ^3-30 \Delta ^2+243 \Delta +90) n\nonumber \\
		&+(\Delta -2) (\Delta -1) \Delta ^2 (\Delta +1)^2 (4 \Delta +3) (4 \Delta +5) (4 \Delta +7) (4 \Delta +9)\\
		\hspace{1.5cm}\nonumber \\
	\mathcal{P}_{3,n}(\Delta)&= 512 (2 \Delta +5) (2 \Delta +7) (2 \Delta +9) n^{12}\nonumber \\
	&+1536 (2 \Delta +5) (2 \Delta +7) (2 \Delta +9) (4 \Delta -1) n^{11}\nonumber \\
	&+128 (2 \Delta +5) (2 \Delta +7) (2 \Delta +9) (264 \Delta ^2-102 \Delta +5) n^{10}\nonumber \\
	&+640 (2 \Delta +5) (2 \Delta +7) (2 \Delta +9) (4 \Delta -1) (44 \Delta ^2-7 \Delta -3) n^9\nonumber \\
	&+96 (2 \Delta +5) (2 \Delta +7) (5280 \Delta ^5+22080 \Delta ^4-8610 \Delta ^3-4790 \Delta ^2-798 \Delta -2931) n^8\nonumber \\
	&+96 (2 \Delta +5) (2 \Delta +7) (4 \Delta -1) (2112 \Delta ^5+9792 \Delta ^4+268 \Delta ^3-5448 \Delta ^2-4628 \Delta -5493) n^7\nonumber \\
	&+8 (2 \Delta +5) (2 \Delta +7) \left(118272 \Delta ^7+556416 \Delta ^6-8736 \Delta ^5-656280 \Delta ^4-661308 \Delta ^3-560400 \Delta ^2 \right. \nonumber \\
	&\quad \left.+371392 \Delta +17415 \right) n^6\nonumber \\
	&+24 (2 \Delta +5) (2 \Delta +7) (4 \Delta -1) \left(8448 \Delta ^7+45504 \Delta ^6+22128 \Delta ^5-79708 \Delta ^4-143680 \Delta ^3-114082 \Delta ^2\right. \nonumber \\
	&\quad \left.+52985 \Delta +27645\right) n^5\nonumber \\
	&+4 (2 \Delta +5) \left( 253440 \Delta ^{10}+2327040 \Delta ^9+5816640 \Delta ^8-1506240 \Delta ^7-22985970 \Delta ^6-33151830 \Delta ^5 \right. \nonumber \\
	&\quad \left.-9079800 \Delta ^4+25792815 \Delta ^3+10370477 \Delta
	^2-446534 \Delta +2131794 \right) n^4\nonumber \\
	&+8 (2 \Delta +5) (4 \Delta -1)\left(14080 \Delta ^{10}+142080 \Delta ^9+423840 \Delta ^8+8160 \Delta ^7-2172753 \Delta ^6-4187481 \Delta ^5 \right. \nonumber \\
	&\quad \left.-1812050 \Delta ^4+3606930 \Delta ^3+3965596 \Delta ^2+1661325
	\Delta +791091 \right) n^3 \nonumber \\
	&+6 (2 \Delta -1) (2 \Delta +5) \left(11264 \Delta ^{11}+125184 \Delta ^{10}+437120 \Delta ^9+118880 \Delta ^8-2771604 \Delta ^7-6808095 \Delta ^6 \right. \nonumber \\
	&\quad \left.-4248981 \Delta ^5+6860955 \Delta ^4+13140919 \Delta
	^3+9496058 \Delta ^2+4002384 \Delta +360360\right) n^2\nonumber \\
	&+2 (2 \Delta +5) (4 \Delta -1) \left(3072 \Delta ^{12}+36096 \Delta ^{11}+132224 \Delta ^{10}-3360 \Delta ^9-1214676 \Delta ^8-2926395 \Delta ^7 \right. \nonumber \\
	&\quad \left.-970776 \Delta ^6+6196080 \Delta ^5+10143424 \Delta
	^4+5128059 \Delta ^3-1542528 \Delta ^2-3028860 \Delta -907200\right) n\nonumber \\
	&+(\Delta -3) (\Delta -2) (\Delta -1) \Delta ^2 (\Delta +1)^2 (\Delta +2)^2 (4 \Delta +5) (4 \Delta +7) (4 \Delta +9) (4 \Delta +11) (4 \Delta +13) (4 \Delta +15)
\end{align}
\tiny
 \begin{align}
 \mathcal{P}_{4,n}(\Delta)&=4096 (2 \Delta +7) (2 \Delta +9) (2 \Delta +11) (2 \Delta +13) n^{16}\nonumber \\
 &+16384 (2 \Delta +7) (2 \Delta +9) (2 \Delta +11) (2 \Delta +13) (4 \Delta -1)
   n^{15}\nonumber \\
   &+4096 (2 \Delta +7) (2 \Delta +9) (2 \Delta +11) (2 \Delta +13) (120 \Delta ^2-46 \Delta +7) n^{14}\nonumber \\
   &+28672 (2 \Delta +7) (2 \Delta +9) (2 \Delta
   +11) (2 \Delta +13) (4 \Delta -1) (20 \Delta ^2-3 \Delta +1) n^{13}\nonumber \\
   &+3584 (2 \Delta +7) (2 \Delta +9) (2 \Delta +11) (4160 \Delta ^5+25792 \Delta ^4-7972 \Delta ^3+492 \Delta ^2-3226 \Delta -4411) n^{12}\nonumber \\
   &+7168 (2 \Delta +7) (2 \Delta +9) (2 \Delta +11) (4 \Delta -1) (1248 \Delta ^5+8320 \Delta ^4+1250 \Delta ^3-1212 \Delta ^2-4345 \Delta -6701) n^{11}\nonumber \\
   &+256 (2 \Delta +7) (2 \Delta +9) (2 \Delta +11) \left( 256256 \Delta ^7+1729728 \Delta ^6+321552 \Delta ^5-936460 \Delta ^4-2427586 \Delta ^3-3105914 \Delta ^2 \right. \nonumber \\
   &\quad \left.+1545816 \Delta -79747\right) n^{10}\nonumber \\
   &+256 (2 \Delta +7) (2 \Delta +9) (2 \Delta +11) (4 \Delta -1) \left(91520 \Delta ^7+681824 \Delta ^6+526504 \Delta ^5-536006 \Delta ^4-2203884 \Delta ^3-3054486 \Delta ^2 \right. \nonumber \\
   &\quad \left.+699565 \Delta +231233\right) n^9\nonumber \\
   &+16 (2 \Delta +7) (2 \Delta +9) \left(13178880 \Delta ^{10}+174839808 \Delta ^9+659459328 \Delta ^8+370120704 \Delta ^7-1507101288 \Delta ^6-4185296472 \Delta ^5\right. \nonumber \\
   &\quad \left.-3717767352 \Delta ^4 +2594242000 \Delta ^3+1067629196 \Delta ^2+296694632 \Delta +503911639\right) n^8\nonumber \\
   &+64 (2 \Delta +7) (2 \Delta +9) (4 \Delta -1) \left(732160 \Delta ^{10}+10396672 \Delta ^9+45167232 \Delta ^8+44861184 \Delta ^7-128566164 \Delta ^6-480365940 \Delta ^5 \right. \nonumber \\
   &\quad \left.-542486796 \Delta ^4+170661206 \Delta ^3+403434322 \Delta ^2+307711299 \Delta +231008470\right) n^7\nonumber \\
   &+224 (2 \Delta +7) (2 \Delta +9) \left(585728 \Delta ^{12}+8639488 \Delta ^{11}+39795712 \Delta ^{10}+40021376 \Delta ^9-176304336 \Delta ^8-651152844 \Delta ^7-674083806 \Delta ^6 \right. \nonumber \\
   &\quad \left.+532826594 \Delta ^5+1165237298 \Delta ^4+880352685 \Delta ^3+360698099 \Delta ^2-378955568 \Delta -17760721\right) n^6 \nonumber \\
   &+224 (2 \Delta +7) (2 \Delta +9) (4 \Delta -1) \left(79872 \Delta ^{12}+1271296 \Delta ^{11}+6677760 \Delta ^{10}+9396800 \Delta ^9-34061064 \Delta ^8-162523434 \Delta ^7-218423448 \Delta ^6 \right. \nonumber \\
   &\quad \left.+107652922 \Delta ^5+541890573 \Delta ^4+598609766 \Delta ^3+245408674 \Delta ^2-202513461 \Delta -83223976\right) n^5\nonumber \\
   &+8 (2 \Delta +7) \left(7454720 \Delta ^{15}+158040064 \Delta ^{14}+1256557568 \Delta ^{13}+4122560512 \Delta ^{12}-544261312 \Delta ^{11}-46507083392 \Delta ^{10}-139137963132 \Delta ^9 \right. \nonumber \\
   &\quad \left.-105682423784 \Delta ^8+287513169874 \Delta ^7+747774586741 \Delta ^6+634799597831 \Delta ^5-73494607795 \Delta ^4-540330812601 \Delta ^3-176220396094 \Delta ^2 \right. \nonumber \\
   &\quad \left.-4065249156 \Delta -37155662874 \right) n^4\nonumber \\
   &+16 (2 \Delta +7) (4 \Delta -1) \left(286720 \Delta ^{15}+6479872 \Delta ^{14}+56057344 \Delta ^{13}+206398976 \Delta ^{12}+5481504 \Delta ^{11}-2738567776 \Delta ^{10}-9579643446 \Delta ^9 \right. \nonumber \\
   &\quad -9234813086 \Delta ^8+23720894549 \Delta ^7+80775271163 \Delta ^6+87715690571 \Delta ^5-1953092747 \Delta ^4-96443198730 \Delta ^3-81493394712 \Delta ^2 \nonumber \\
   &\quad \left.-33710278446 \Delta -13620018456\right) n^3\nonumber \\
   &+8 (2 \Delta +7) (2 \Delta -1) \left(245760 \Delta ^{16}+5926912 \Delta ^{15}+55634944 \Delta ^{14}+227535616 \Delta ^{13}+40873280 \Delta ^{12}-3507911344 \Delta ^{11}-14104347484 \Delta ^{10} \right. \nonumber \\
   &\quad \left.-16399401415 \Delta ^9+41696678246 \Delta ^8+175336044542 \Delta ^7+232465979473 \Delta ^6+14198330860 \Delta ^5-344274915821 \Delta ^4-451122478755 \Delta ^3 \right. \nonumber \\
   &\quad \left.-288164187918 \Delta ^2-107494932816 \Delta -9428098680\right) n^2\nonumber \\
   &+8 (2 \Delta +7) (4 \Delta -1) \left(16384 \Delta ^{17}+413696 \Delta ^{16}+4072448 \Delta ^{15}+17113344 \Delta ^{14}-3852352 \Delta ^{13}-348478480 \Delta ^{12}-1390768244 \Delta ^{11} \right. \nonumber \\
   &\quad \left.-1287252149 \Delta ^{10}+6772064077 \Delta ^9+24985583487 \Delta ^8+28270657594 \Delta ^7-20460756119 \Delta ^6-94962146759 \Delta ^5-105214119603 \Delta ^4 \right. \nonumber \\
   &\quad \left.-36550274148 \Delta ^3+24596517024 \Delta ^2+29350981800 \Delta +7858620000\right) n \nonumber \\
   &+(\Delta -4) (\Delta -3) (\Delta -2) (\Delta -1) \Delta ^2 (\Delta +1)^2 (\Delta +2)^2 (\Delta +3)^2 (4 \Delta +7) (4
   \Delta +9) (4 \Delta +11) (4 \Delta +13) (4 \Delta +15) (4 \Delta +17) (4 \Delta +19) (4 \Delta +21)
\end{align}

\newpage
\bibliographystyle{nb}
\bibliography{Mellin1d_v8}

\end{document}